\documentclass[12pt, a4paper,onecolumn,notlepage] {article}
\pdfoutput=1
\usepackage{multirow,color,amsmath}
\usepackage{cite}
\usepackage{hyperref}
 \usepackage{graphicx}
 \usepackage{caption}
\usepackage{subcaption}
\usepackage{longtable}
\usepackage{ulem}

\begin{document}
\renewcommand{\thefootnote}{\fnsymbol{footnote}}
\setcounter{footnote}{0}
\title{\textbf{Theories and Practice of Agent based Modeling: Some practical Implications for Economic Planners} }
\author{Hossein Sabzian \\
\thanks {Iran University of Science and Technology.\uline{\textbf{hossein\_sabzian@iust.ac.ir}}}\\
\and Mohammad Ali Shafia\\
\thanks {Iran University of Science and Technology.\uline{\textbf{omidshafia@ iust.ac.ir}}}\\
\and Ali Maleki\\
\thanks {Sharif University of Technology.\uline{\textbf{a.maleki@sharif.edu}}}
\and Seyeed Mostapha Seyeed Hashemi\\
\thanks {Tadbir Economic Development Group.\uline{\textbf{smseyeedhashemi@yahoo.com}}}
\and Ali Baghaei\\
\thanks {Tadbir Economic Development Group.\uline{\textbf{Alibaghaei@yahoo.com}}}
\and Hossein Gharib\\
\thanks {Tadbir Consulting Group.\uline{\textbf{gharib1337@gmail.com}}}
}
\maketitle

\renewcommand{\thefootnote}{\arabic{footnote}}
\begin{abstract}
Nowadays, we are surrounded by a large number of complex phenomena such as virus epidemic, rumor spreading, social norms formation, emergence of new technologies, rise of new economic trends and disruption of traditional businesses. To deal with such phenomena, social scientists often apply reductionism approach where they reduce such phenomena to some lower-lever variables and model the relationships among them through a scheme of equations (e.g.  Partial differential equations and ordinary differential equations). This reductionism approach which is often called equation based modeling (EBM) has some fundamental weaknesses in dealing with real –world complex systems, for example in modeling how a housing bubble arises from a housing market, the whole market is reduced into some factors (i.e. economic agents) with unbounded rationality and often perfect information, and the model built from the relationships among such factors is used to explain the housing bubble while adaptability and the evolutionary nature of all engaged economic agents along with network effects go unaddressed. In tackling deficiencies of reductionism approach, in the past two decades, the Complex Adaptive System (CAS) framework has been found very influential. In contrast to reductionism approach, under this framework, the socio-economic phenomena such as housing bubbles are studied in an organic manner where the economic agents are supposed to be both boundedly rational and adaptive. According to CAS framework, the socio-economic aggregates such as housing bubbles emerge out of the ways agents of a socio-economic system interact and decide. As the most powerful methodology of CAS modeling, Agent-based modeling (ABM) has gained a growing application among academicians and practitioners. ABMs show how simple behavioral rules of agents and local interactions among them at micro-scale can generate surprisingly complex patterns at macro-scale. Despite a growing number of ABM publications, those researchers unfamiliar with this methodology have to study a number of works to understand (1) the why and what of ABMs and (2) the ways they are rigorously developed. Therefore, the major focus of this paper is to help social sciences researchers get a big picture of ABMs and know how to develop them both systematically and rigorously.
\end{abstract}

 \section{Introduction} \label{section.Intro}
 A large number of social phenomena such as cultural changes, cooperation formation, innovation, norm formation, technology diffusion, and even evolution of states happen not just due to separate choices by constituent individuals but mainly because of dynamic interactions among them over time. As a matter of fact, such phenomena have a nature entirely different from their constituents. Modeling the formation of these collective phenomena has been a great target for mainstream socio-economic modeling approach but it has not captured it sufficiently. This mainstream modeling approach which often called equation-based modeling (EBM) has been frequently used in different disciplines of social sciences. However, EBMs lack a needed functionality in explaining how the interactions among micro-components of a system can lead to an interestingly different macro-behavior for that system. In fact, they perform very poorly in modeling the emergent properties of real-life systems, namely how a whole arises from the interactions among its simpler and lower-level parts so that it exhibits properties that its simpler and lower-level parts can never exhibit. For tackling such a limitation, the agent based models (ABMs)  have been developed. An ABM is a kind of computational model which explores systems of multiple interacting agents which are spatially situated and evolve over the time. ABMs are highly effective in explaining how complex patterns emerge from micro-level rules during a period of time. In contrast to EBMs that are based on deductive reasoning, ABMs properly work not only as an inductive reasoning technique where a conclusion is formed from a series of observations but also as a pure form of abductive reasoning where the best explanation for the phenomena under study is inferred via simulation. ABMs have become a major modeling trend in a large number of domains ranging from spread of epidemics\cite{Situngkir2004} and the threat of bio-warfare\cite{Caplat2008} to formation of norms\cite{Axelrod1986}, supply chain optimization\cite{VanDykeParunak1998,Jetly2014,Swaminathan1998} and collaboration in project teams\cite{Son2010}. In contrast to EBMs which majorly focus on relationship among macro-variables of a system in top-down manner, ABMs try to model how local and predictable interactions among micro-components of a system can generate a complex system-level behavior\cite{Macy2002}. ABM methodology is rooted in complexity theory and network science. In terms of complexity theory, ABMs are developed to explain how simple rules generate complex emergence (i.e. a process model) and in terms of network science ABMs are used to analyze the pattern that arise from agents’ interactions over the time (i.e. a pattern model)\cite{wilensky2015introduction}.In this paper, we want to explore ABMs systematically and show their great potentiality for modeling a large number of real world problems (\textrm{with a special focus on social problems}) that contemporary methods cannot model properly.\\
The rest of this paper is organized as follows: the \hyperref [section.whyandwhat]{section \ref*{section.whyandwhat}} deals with why and what of ABMs.The unique characteristics of ABMs in comparison to EBMs are discussed in \hyperref [section.unique]{section \ref*{section.unique}}. \hyperref [section.main]{section \ref*{section.main}} is concerned with main uses of ABMs.ABM building blocks are discussed in \hyperref [section.Buildingblocks]{section \ref*{section.Buildingblocks}}. ABM development process is unraveled in  \hyperref [section.development]{section \ref*{section.development}}. Some critical considerations are offered in  \hyperref [section.considerations]{section \ref*{section.considerations}}.Two economics-related applications of ABMs are presented in \hyperref [section.simulations]{section \ref*{section.simulations}} and a conclusion is provided  \hyperref [section.conclusion]{section \ref*{section.conclusion}}.

 \section{Why and What of ABM}\label{section.whyandwhat}
 We are living in complex world which itself includes an unlimited number of complexities ranging from highly micro-level complexities such as interacting atoms to highly macro-level ones such as nations. With an eye to socio-economic organizations like banks, insurance companies, hospitals and automobile producers, it becomes clear that all of these organizations are in turn a type of complex system so that each of them owns a distinguished whole (or ensemble) beyond its constituent parts (or components). Complex systems should be considered different from complicated systems. Actually, a complex system includes multiple interacting components forming a whole irreducible to its parts, therefore, it doesn’t lend itself to divide-and-conquer logic while a complicated system is composed of multiple related components forming a while reducible to its part and can be understood by divide-and-conquer logic. When a complex system is studied, the uncertainty of its outcomes never decreases to zero but as soon as a complicated system is analyzed and understood, the certainty of its outcomes increases to a large degree\cite{Snyder2013}.
\begin{figure} 
    \centering
    \begin{subfigure}[b]{0.7\textwidth}
        \includegraphics[width=\textwidth, , trim = 5px 5px 5px 5px,  clip = true]{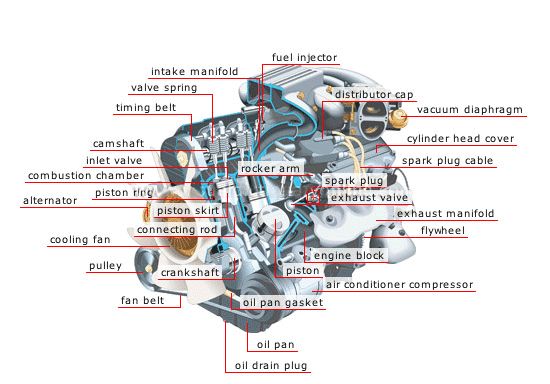}
        \caption{An engine\footnote{https://supamac.co.za/looking-inside-your-car-engine/} }
        \label{Engine}
    \end{subfigure}
 ~
    \begin{subfigure}[b]{0.7\textwidth}
        \includegraphics[width=\textwidth, , trim = 0 0 0 1px,  clip = true]{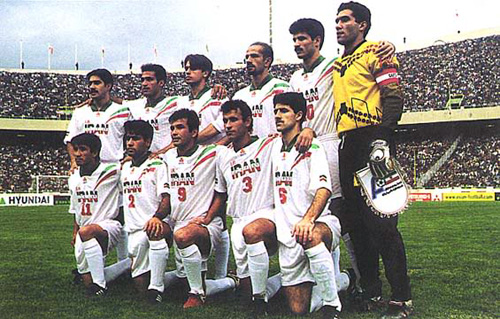}
        \caption{A Team as a complex system\footnote{http://www.parstimes.com/soccer/players.html}}
        \label{Team}
    \end{subfigure}
    ~
    \caption{Two different systems}\label{systems}

\end{figure} 
One example for illustrating the difference between a complex system and a complicated one is Figure\ref{Engine} and Figure\ref{Team}. A car engine is assembled by several number of parts. When it is well understood by a team of experts, it can be decomposed and integrated over and over again without losing any of its expected functionalities.A car engine is assembled by several number of parts. When it is well understood by a team of experts, it can be decomposed and integrated over and over again without losing any of its expected functionalities. In contrast, a team including a number of interacting persons can show a surprisingly unexpected performance even the experts disarrange it from its initial conditions and rearrange it completely the same as its prior initial conditions.\footnote{It can be inferred that what Merton calls “unintended consequences” can just be observed in complex systems such as society\cite{Merton1936}}Complexity theory (CT) is an interdisciplinary filed of studying complex systems ranging from biophysical complex systems such as molecules and organs to socio-economic complex systems such as small firms and multi-national corporations. According to CT, complex systems which absorb information from surrounding environment and accumulate knowledge that can help action are usually called complex adaptive systems (CASs). A CAS represents the notion of a system where “The whole is more than the parts”. Actually, these are systems where multiple and perhaps very simple parts interact in a nonlinear and non-trivial manner to give rise to global often unpredictable behaviors observable and discoverable at a higher level of abstraction\cite{Holland2002}. A number of fundamental characteristics of CASs are listed in Table \ref{table1}.\\
In the domain of CASs modeling methodologies\footnote{CAS modeling methodologies have been comprehensively discussed in \cite{Niazi2011} } , ABMs as a micro-scale computational models\footnote{Microscale models form a broad class of computational models that simulate fine-scale details, in contrast with macroscale models, which amalgamate details into select categories. Microscale and macroscale models can be used together to understand different aspects of the same problem \cite{Gustafsson2007,Gustafsson2010}} have shown a much better performance than equation-based models (EBMs) such as analytical models and statistical modeling methods\cite{EliotR.Smith2007,niazi2017towards,Sun2005,VanDykeParunak1998,wilensky2015introduction}. Developed from the fields of complexity, cybernetics, cellular automata and computer science, ABMs have gained lots of popularity in the 1990s and shows a growing migration not only from equation based models (EBMs) such as econometric models, analytical models and statistical modeling techniques but also from the more classical simulation approaches such as the discrete-event simulation\cite{Heath2009,Siebers2008,Siebers2010}. ABMs have a wide range of application domains ranging from biological systems\cite{Caplat2008,Situngkir2004} to engineered ones \cite{Olfati-Saber2004}. The primary reason widely held by ABM practitioners is its very high strength in modeling complex adaptive systems (CAS) in comparison with other modeling methods.
\begin{longtable}{| p{.25\textwidth} | p{.75\textwidth} |}
\caption{Fundamental characteristics of CASs}
\label{table1}\\
  \hline
  Characteristics & Descriptions\\
    \hline
 Multiplicity and heterogeneity of constituent components & It is composed of a number of components usually called” agent” \cite{wilensky2015introduction,Holland2002}.These agents can be very heterogeneous.  \\
    \hline
 Non-linear interactions &Its agents interact with each other in a non-linear (non-additive) way \cite{wilensky2015introduction,Holland2002}.  \\
    \hline
  Learnability and adaptability & Its agents can adapt or learn\cite{Holland2002} so agents can experience and accumulate knowledge.\\
    \hline
Non-ergodicity &It is non-ergodic\cite{Kauffman2000,Moss2008}.Therefore, it is highly sensitive to initial conditions. \\
   \hline
  Self-organization & It self-organizes and its control is intensely distributed among its agents\cite{Chan2001, wilensky2015introduction}. \\
    \hline
  Emergence & It exhibits emergence \cite{wilensky2015introduction,Holland2002,Chan2001}. It means, from the interactions of individual agents arises a global pattern or an aggregate behavior which is characteristically novel and irreducible to behavior(s) of agent(s). \\
    \hline
 Co-evolution &Its agents can co-evolve and change the system’s behavior gradually\cite{Kauffman2000}.  \\
  \hline
 Far from equilibrium &It shows “far from equilibrium” phenomenon\cite{nicolis1989exploring}. Isolated systems have a high tendency towards equilibrium and this will cause them to die. The “far from equilibrium” phenomenon shows how systems that are forced to explore possibilities space will create different structures and novel patterns of relationships\cite{Chan2001, wilensky2015introduction}.\footnotemark[1]\\
  \hline
 Time asymmetry and irreversibility & It is time asymmetric and irreversible. One characteristic of a CAS is time asymmetry. Asymmetry in time occurs when a system passes a bifurcation point, a pivotal or decisional point where an option is taken over another or others, leading to time irreversibility. Irreversibility means that the system cannot be run backwards— rewound or reversed—so as to reach its exact initial conditions. Systems which, when run in reverse, do not necessarily or typically return to their original state are said to be asymmetric in time\cite{Prigogine1997}, and asymmetry in time is important in testing for a complex adaptive system. If system-time is symmetric in both directions, then it is reversible, and it is not a CAS but a deterministic system. Complex adaptive systems are asymmetric in time, irreversible and nondeterministic. So, in a CAS one can neither predict nor “retrodict,” even with infinite information on initial conditions, because the system “chooses” its forward path. Its “choice” is indeterminate, a function of statistical probability rather than certainty\cite{Rogers2005}.\\
 \hline
 Distributed control & The behavior of a CAS is not controlled by a centralized mechanism, rather, it is completely distributed among its constituent parts.  The interactions of these constituent parts cause a CAS to exhibit a coherent macro-level behavior\cite{Chan2001}.\\
 \hline
\end{longtable}
\footnotemark[1] In thermodynamics, systems that does not have any exchange of energy and matter with their surrounding environment are called “isolated systems”. Such systems have a tendency to evolve towards equilibrium. But, our surrounding is enriched by phenomena arising from conflations far from equilibrium. Some examples can be turbulences, fractals and even life itself \cite{Jaeger2010}.
\\
The philosophy of agent-based modeling comes directly from the idea that a CAS can be effectively modeled and explained by creating agents and environment, characterizing behavioral rules of agents, and specifying interactions among them \cite{wilensky2015introduction}. Modeling a CAS needs a specific type of methodology. EBMs such as statistical modeling techniques or PDEs lack a needed functionality for this purpose because they just decompose a system into its main parts and model the relationship among them (a top-down approach) while neglecting the fact that the system itself is entity beyond its constituent parts and it needs to be analyzed as an emergence of its constituent parts (a bottom-up approach).

 \section{Unique characteristics of ABMs in comparison with EBMs} \label{section.unique}
EBM and ABM have stemmed from two distinct epistemological frameworks. The former is grounded on reductionism approaches such as neoclassical economic theories (NET) where the issues such as unbounded rationality, perfect information, deductive reasoning and low-rate heterogeneity are discussed while the latter is built upon complexity theory (CT)where the issues such as bounded rationality, information asymmetry, network interaction, emergence and inductive reasoning are taken into consideration \cite{Al-suwailem2008}. This has made ABMs specifically advantaged in modeling CASs. Some of these advantages can be summarized as Table\ref{table2}
\begin{longtable}{| p{.25\textwidth} | p{.75\textwidth} |}
\caption{Major advantages of ABM over EBM} 
\label{table2}\\
  \hline
Advantage & Description\\
    \hline
Bounded rationality & The environment in which agents interact is highly complex and unbounded rationality is not a viable assumption for it \cite{Al-suwailem2008 ,wilensky2015introduction}, agents have limited possibilities not only for receiving information but also for its processing. AB modelers contend that socio-economic systems have an inherently non-stationary nature, due to continuous novelty (e.g., new patterns of aggregated behavior) endogenously introduced by the agents themselves\cite{Windrum2007}. Therefore, it is extremely difficult for agents to learn and adapt in such a turbulent and endogenously changing environments. On this basis, ABM researchers argue that assumption of unbounded rationality is an unsuitable for modelling real world systems and agents should not only have bounded rationality but also adapt their expectations in different periods of time.\\
    \hline
Exhibition of emergence &Since ABMs can model how micro-dynamics result in a high-level macro-dynamic they can be used as the best method for exhibiting emergent properties. On this basis, ABM does not require knowledge of the aggregate phenomena, in fact, researchers do not need to know what global pattern results from the individual behavior. When modeling an outcome variable with EBM, you need to have a good understanding of the aggregate behavior and then test out your hypothesis against the aggregate output \cite{wilensky2015introduction}.  \\
    \hline
 Bottom-up perspective& A macro-system is an outcome of the way its sub-systems interact so the properties of macro-dynamics can only be properly understood as the outcome of micro-dynamics involving basic entities/ agents\cite{Tesfatsion2002}. This contrasts with the top-down nature of traditional neoclassical models, where the bottom level typically comprises a representative individual and is constrained by strong consistency requirements associated with equilibrium and unbounded rationality\cite{EliotR.Smith2007, Macy2002}. Conversely, AB models describe strongly heterogeneous agents living in complex systems that evolve through time \cite{Kirman1997, Kirman2005}. Therefore, aggregate properties are interpreted as emerging out of repeated interactions among simple entities rather than from the consistency requirements of rationality and equilibrium imposed by the modeler \cite{Dosi1994}.\\
    \hline
Heterogeneity and discrete nature &An ABM can nicely model a heterogeneous population, whereas equational models typically must make assumptions of homogeneity. In many models, most notably in social science models, heterogeneity plays a key role. Furthermore, when you model individuals, the interactions and results are typically discrete and not continuous. Continuous models do not always map well onto real-world situations \cite{wilensky2015introduction}. \\
   \hline
Networked interactions: &Interactions among economic agents in AB models are direct and inherently non-linear \cite{Fagiolo1998,Silverberg1988} . Agents interact directly because current decisions directly depend, through adaptive expectations, on the past choices made by other agents in the population (i.e. a widespread presence of externalities). These may contain structures, such as subgroups of agents or local networks. In such structures, members of the population are in some sense closer to certain individuals in the socio-economic space than others. These interaction structures may themselves endogenously change over time, since agents can strategically decide with whom to interact according to the expected payoffs. When combined with heterogeneity and bounded rationality, it is likely that aggregation processes are non-trivial and, sometimes, generate the emergence of structurally new objects \cite{Lane1993a,Lane1993}. \\
    \hline
 Comprehensiveness:e & It exhibits emergence \cite{wilensky2015introduction,Holland2002,Chan2001}. Results generated by ABMs are more detailed than those generated by EBMs. ABMs can provide both individual and aggregate level detail at the same time. Since ABMs operate by modeling each individual and their decisions, it is possible to examine the history and life of any one individual in the model, or aggregate individuals and observe the overall results.  This “bottom-up” approach of ABMs is often in contrast with the “top-down” approach of many EBMs, which tell you only how the aggregate system is behaving and do not tell you anything about individuals. Many EBMs assume that one aspect of the model directly influences, or causes, another aspect of the model, while ABMs allow indirect causation via emergence to have a larger effect on the model outcomes \cite{wilensky2015introduction}. \\
    \hline
Randomness and indeterminacy & One important feature of agent-based modeling, and of computational modeling in general, is that it is easy to incorporate randomness into your models. Many equation-based models and other modeling forms require that each decision in the model be made deterministically. In agent-based models this is not the case; instead, the decisions can be made based on a probability \cite{Siebers2008,wilensky2015introduction}.  \\
 \hline
\end{longtable}
With regard to Table \ref{table2}, it makes sense that social structures such as teams, organizations, governments and nations or even galaxial systems are few examples of CASs each of which can exhibits a number of emergent properties. For instance, organizations are a type of CAS out of which phenomena such as cooperation, aggregation of core competencies or even the ways employees interactively reinforce or weaken organizational routines emerge\cite{Wall2016}. In a wider economic system, macro-level phenomena such as inflation, stagflation, stock markets dynamics and economic inequality are aggregates (complex problems) emerging out of the economic systems. In recent years, the literature about complexity economics has been developed in so many areas including evolutionary models inspired by Nelson and Winter \cite{Nelson1982}and Hodgson \cite{Hodgson1998}, Brock and Durlauf’s study of social interaction\cite{Brock2001} , study of firm size by Axtell\cite{Axtell2001}, Alan Kirman and his colleagues models of financial markets\cite{Kirman2005} and the agent-based simulation of general equilibrium\cite{Gintis2006a,Gintis2006}.\footnote{For a comprehensive overview of computational methods in complexity economics, look at \cite{Amman1996,Tesfatsion2002}}\\
However, regarding complex nature of real world, it goes clear that EBMs (such as constrained optimization models used in econometrics) cannot capture the behavior of complex adaptive systems. This is an essential departure from the presumptions existing in conventional economic theories. Such systems should be analyzed ‘in’ time and this limits the way that mathematics can be used. Standard economic theory includes the applying of an ahistorical body of logical clauses to display attitudes perceived in the historical domain. In opposition, complex adaptive system theory copes directly with the fundamental principles that rule the behavior of systems in history. Therefore, it can be said that thinking about the economy and its sub-components as complex adaptive systems can allow us to evade such scientific impasses .In economic thought, Schumpeter’s contributions toward the process of “creative destruction” conform to complex adaptive systems theory\cite{Foster2001}.
 However, the core idea of Agent-Based Modeling is rooted in the fact that  a CAS can be productively modeled with agents, an environment, and the rules of interactions among them. An agent is an autonomous entity with particular properties, behaviors, and even goals.  The environment is a landscape over which agents have interactions and can be spatial, network-based, or mixture of them. The interactions can be non-linear and quite complex. Agents can have interaction with other agents or with the environment and they can not only change their interaction rules but also can change the strategies used to decide what behavior to do at a particular time \cite{wilensky2015introduction}. So, ABMs can be considered as a revolutionary methodology for modeling and simulating systems (i.e. real-world CASs) that are tremendously difficult and often impossible to be studied by EBMs\cite{Bankes2002}.

 \section{Main uses of ABMs} \label{section.main}
ABMs can be used in description and explanation. Like all models, an AMB is a simplification of a real world system which doesn’t entail all of its aspects so it is distinguishable from real world system and can help its understanding. The exploratory nature of ABM indicates that they can be used to pinpoint the essential mechanisms underlying the phenomena under study. a subject matter expert can use an AMB as a proof that his or her hypothesized mechanisms sufficiently account for the aggregate behavior under study \cite{wilensky2015introduction}. Explanation is strongly believed to be a major function of ABMs because it helps understand how simple rules generate complex structures. ABMs’ explanatory power is highly generative, especially in social sciences due to the fact that it explains which macro-structures such as epidemic dynamics or social evolutions emerge in population of heterogeneous agents that interact locally and in non-trivial way under a set of tenable behavioral rules \cite{Epstein2008}.\\
ABMs facilitate the experimentation process \cite{Leal2017}. They can be run repeatedly to discern variations in their dynamics and in their outputs \cite{wilensky2015introduction}. Some models show a very little variations during several runs. Some have a path-dependency nature \cite{Brown2005} and some exhibit tremendous variations from run to run. Through experimentation, system modelers get informed of how input parameters affect model’s outputs. Therefore, they can make various scenarios for achieving the targeted behavior.\\
ABMs are sometimes used for prediction purposes. Subject matter experts frequently use models to get a picture about possible future states. Like every model the quality of ABMs’ prediction relies on the accuracy of its input parameters and since society is a complex system with an unspecified degree of uncertainty and very high sensitivity to small-scale events, no prediction can be deemed as absolutely right \cite{Moss2008,wilensky2015introduction}. Prediction differs from description where the modeler describes the past or present states of the system, for example when a modeler describes what changes first occurred in the system. Moreover, prediction is also district from explanation, for example Plate tectonics definitely explains earthquakes, but does not help us to predict the time and place of their occurrence or evolution is commonly accepted as explaining speciation, but it is impossible to predict next year's flu strain \cite{Epstein2008}.Nonetheless, when subject matter experts claim to have used ABMs for purpose of prediction, they actually use ABMs either for description or explanation \cite{wilensky2015introduction}.\\
ABMs has a high functionality for education and analysis \cite{Blikstein2009,Wilensky2006,Sengupta2009}. Educators can develop models for people that they have never seen before. For example, educators can model some examples of mutualism between individuals of different species when both individuals benefit\footnote{An interesting example can be the mutualism between a goby and a shrimp. The shrimp digs a burrow in the sand and cleans it up where both species can live.  Since the shrimp is almost blind, it has a high vulnerability to predators outside the burrow. When the shrimp is under dangerous conditions the goby goes over to warn the shrimp by touching it with its tail. This causes  both the shrimp and goby quickly back into the burrow \cite{Helfman2009}}Moreover, models can simulate a system that may not be readily available from real-world observations, therefor they can be very thought-provoking and enable learners to go beyond their observations and conduct experiments just like scientists.\\
When a subject matter expert is going to gain a deeper understanding of a phenomena about which there is not enough theory, thought experiment can be very useful. Though experiment is another suitable area for ABMs. This type of experiment is done to achieve its purpose without benefit of execution \cite{Sorensen1998}.Thought experiment is conducted when the real-world experiments are neither affordable nor possible to execute \cite{Rangoni2014}. It has a wide application in social and natural sciences. Through this method, researchers can get aware of the logical consequences of their hypotheses. For example, what will happen if a half of a company’s staff suddenly leave it? ABMs can be very useful in thought experiments especially when people want to deal with complex systems such organization and society. Such systems are far from a real-word laboratory where it is possible to control some variable (as control group) and measure the effect of test on other variables (as treatment group). As a matter of fact, in such systems, there are numerous causal factors that are mainly interdependent over which we have on or a very limited control\cite{Savona2005}. So real-world experiments can rarely be executed in such systems. This has led researchers of social fields to utilize the potential of though experiment in simulating the consequences of their hypothesized mechanism.\\

  \section{ABM Building Blocks} \label{section.Buildingblocks}
\begin{center}
\begin{figure}
\centering
    \includegraphics[scale = 0.2 ,trim = {7cm 3cm 3cm 8cm}, clip = true  ]{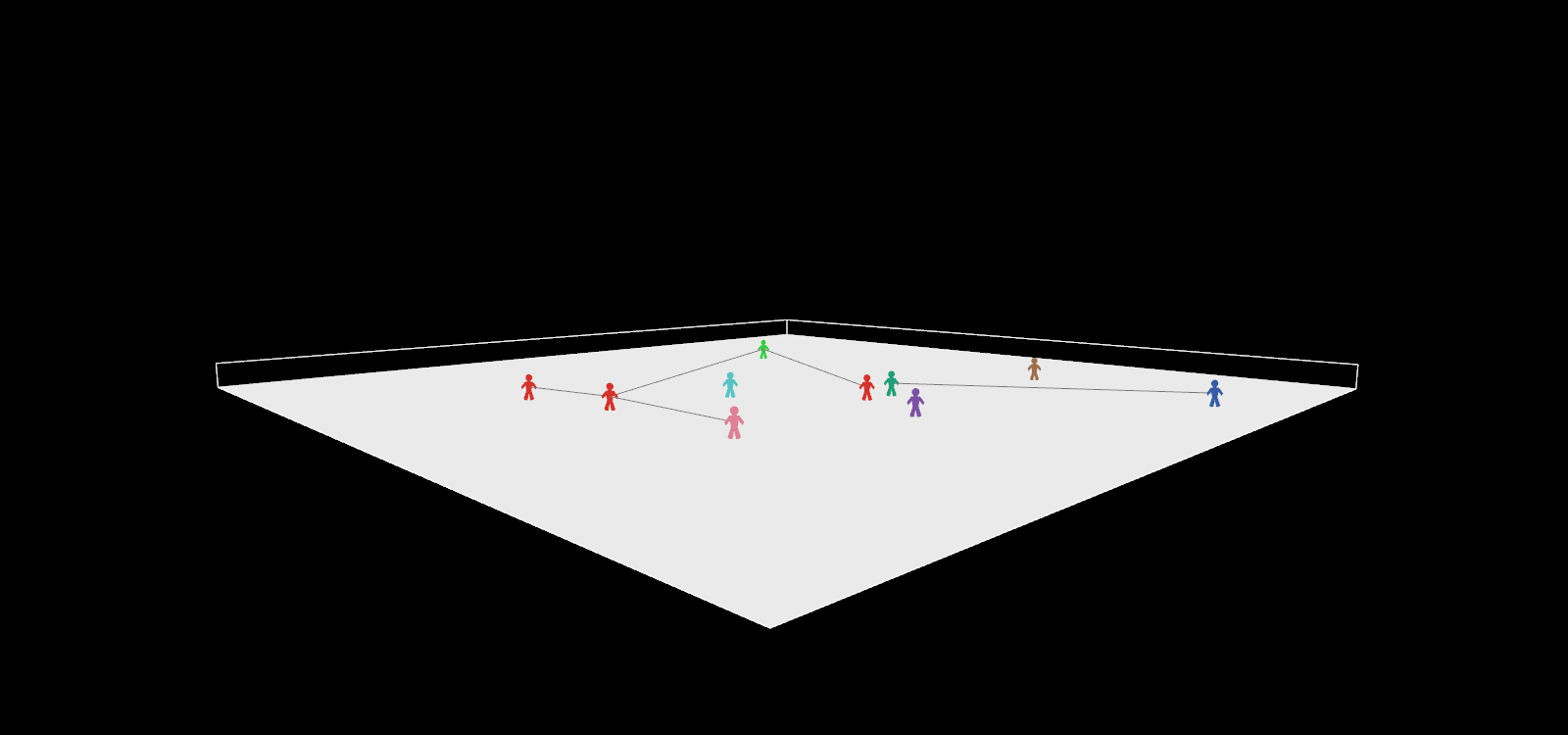}
    \caption{Building Blocks of ABM}
    \label{bblocks}
\end{figure}
\end{center}

\begin{center}
\begin{figure}
\centering
    \includegraphics[scale = 0.7 ,trim = {20cm 8cm 20cm 9cm}, clip = true  ]{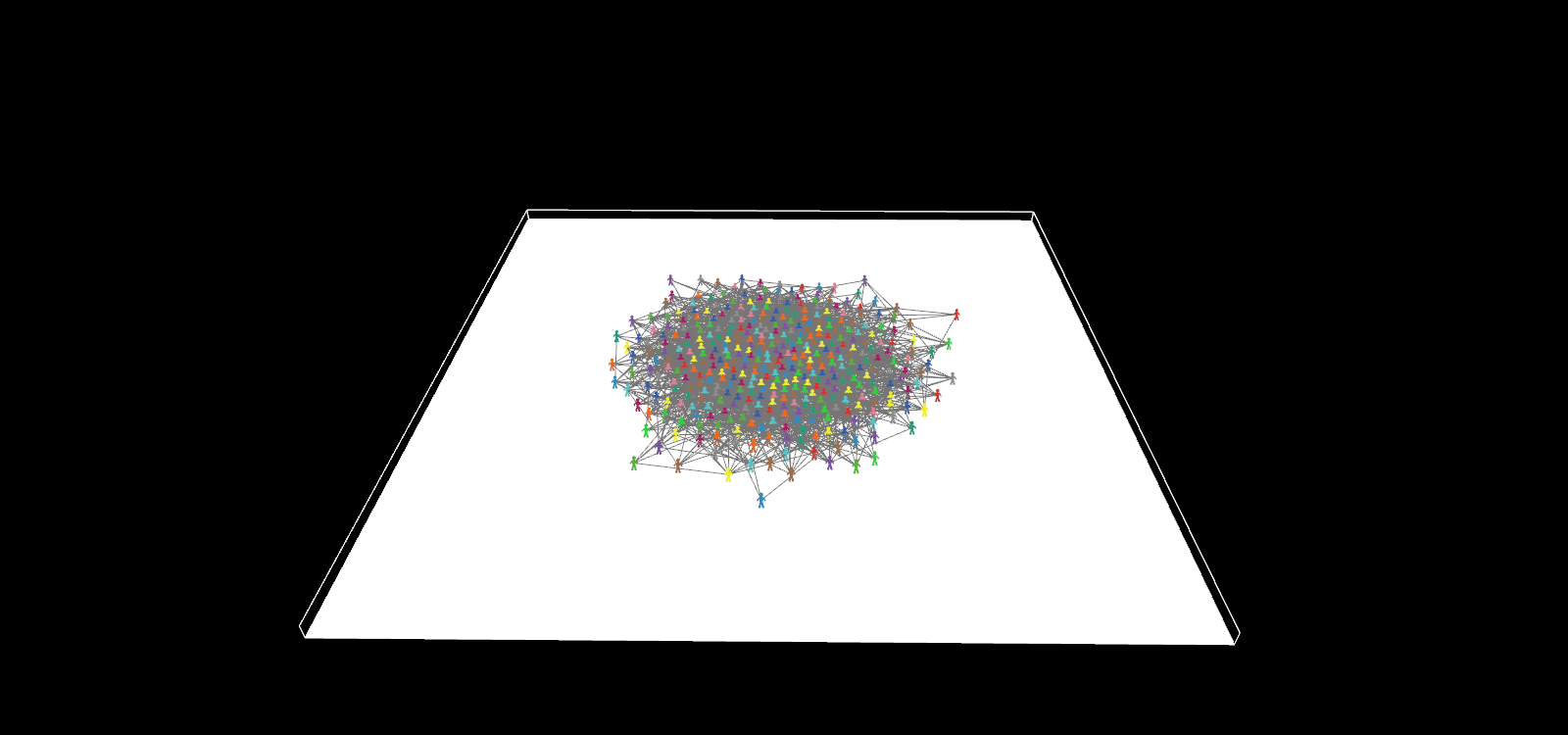}
    \caption{A random network}
    \label{Random}
\end{figure}
\end{center}

\begin{center}
\begin{figure}
\centering
    \includegraphics[scale = 0.4 ,trim = {3cm 3cm 4cm 3cm}, clip = true  ]{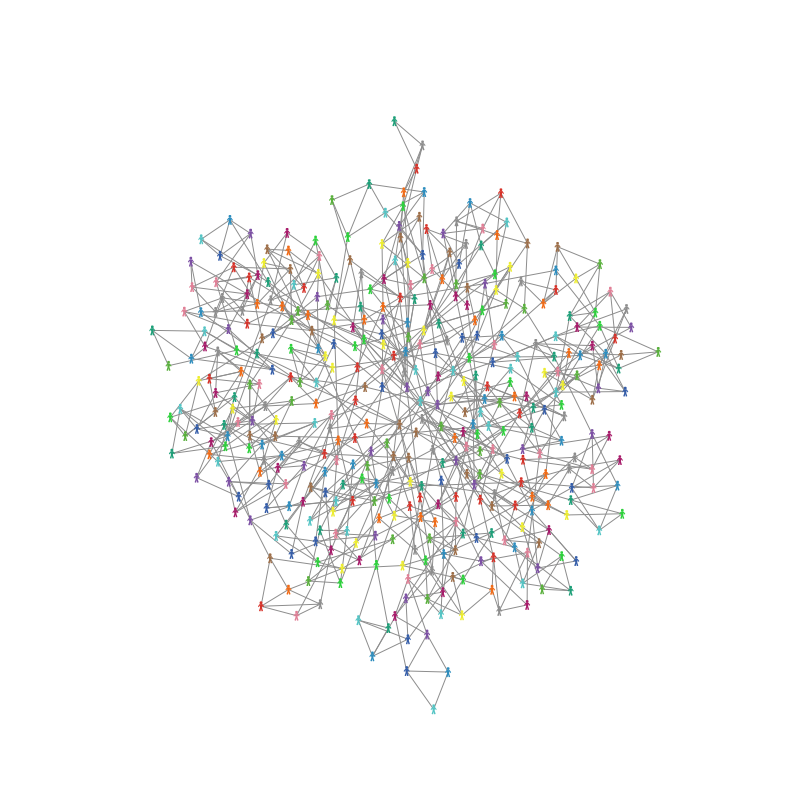}
    \caption{Small-world network}
    \label{SW}
\end{figure}
\end{center}

\begin{center}
\begin{figure}
\centering
    \includegraphics[scale = 0.4 ,trim = {3cm 3cm 3cm 3cm}, clip = true  ]{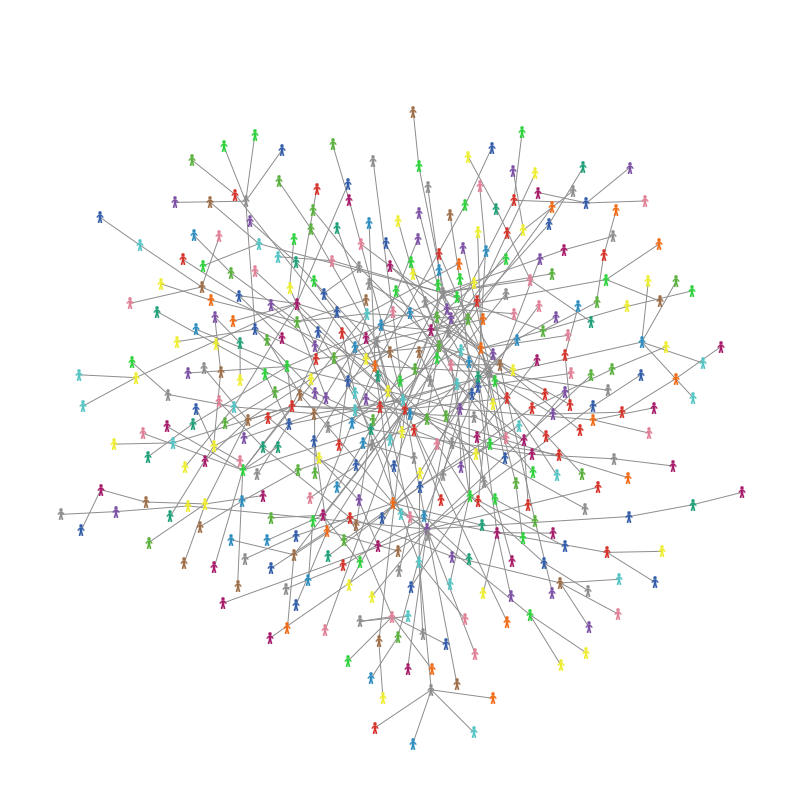}
    \caption{Scale-free preferential attachment network}
    \label{PA}
\end{figure}
\end{center}

ABMs include three building blocks of (1) agents, (2) environment and (3) interactions \cite{Epstein1997, niazi2017towards,wilensky2015introduction}. As the first building block of ABMs, agents are the basic computational units of agent based models. They are defined by two main aspects of (1) properties and (2) behaviors (or actions). Agent’s properties are internal or external states that can be changed by its behaviors (actions).  Suppose you want to model an economic system including individual human agents. Some properties for these agents can be status of employment, income level, number of bank account and age or even if necessary blood type! Actions of such agents can be searching for a job, opening a bank account, taking a loan and so on. As it is sensible, actions affect properties, for example, when a person opens a new bank account, the number of his or her bank accounts increases. Or when a person finds a job, his or her status of employment is changed and subsequently his or her income level is positively influenced. As the first building block of any ABM, agents are in three specific types of mobile agents, stationary agents and connecting agents. Mobile agents have the capability of movement for example a human is a type of a mobile agent. Stationary agents are those static agents that have no moving capability. For example, an organization or in wider sense, an environment are types of stationary agents\footnote{According to Netlogo programming language, an environment is built of several static agents called “patch” and all agents are on the environment. ABM programming languages and tools will be discussed in  \hyperref [section.Development.Programming]{subsection \ref*{section.Development.Programming}}}. Connecting agents are those agents that connect agents together. One clear example of this can be “links” among agents (\ref{bblocks}). Additionally, in modeling agents, two major factors have to be taken into consideration, the first factor is about the granularity (grain-size) of agents. For example, when you want to model an economic system, you chose to model the individual actors or prefer to model institutions. The second important factor deals with the cognitive level of agents. In fact, how much is the capability of agents to observe (and sense) the surrounding world and make decisions?\\
According to cognitive level of agents, they can be classified into four types of (1) reflective or myopic agents, (2) utility-based agents, (3) goal-based agents, (4) adaptive agents \cite{wilensky2015introduction}. The reflective agents are very simple if-then agents so that if they face situation A, they immediately do action B. Utility-based agents are very similar to reflective ones but there is a utility function that they do want to maximize it under all conditions. Goal-based agents are more advanced form of utility-based function so that they have a goal that dictates their actions. The most advanced form of agents are adaptive agents. They have enough cognitive capabilities to change their actions in similar conditions based on prior experience. Namely, if they do action A in situation B and lose some payoffs, when they face situation B again, they don’t do action A according to their prior experiences.\footnote{For a more comprehensive study of agent cognition, look at \cite{Russell2016}}\\
As the second building blocks of ABMs, the environment is composed of all conditions surrounding the agents as they interact within the model. In other words, the environment is where an artificial social life unfolds \cite{Epstein1997}. Environments can come into three different major forms of (1) spatial environment, (2) networked environment and (3) mixed environment. The spatial environment is often a discrete environment including several discrete points\footnote{It can also be continuous, see \cite{wilensky2015introduction}} . The most common form of spatial environment is lattice structure which can be two or three dimensional (as Figure \ref{bblocks}). In spatial environments, when agent A (here the purple agent) reaches a border on the far right side of the environment (i.e., the world) and wants to go farther right, boundary conditions of the environment come to play. The topology of an environment deals with such boundary conditions. For a spatial lattice structure such as Figure \ref{bblocks}, there can be three types of topologies. The first type is a toroidal topology where agent A reappears in the far left side of the lattice. The second type is   bounded topology where agent A cannot move farther right and finally, the third type is infinite plane topology where agent A can keep going right for ever \cite{wilensky2015introduction}. In real world situations, such as socio-economic settings, agents have more networked interactions than spatial (geographical) interactions. In two different stock markets, a rumor spreads through the individual agents of a network. Therefor an environment can be in a network form where the mobile agents are “nodes” and the connections among them are “links”. There are several types of networks that three of them are widely used which are  “random networks” \cite{Erdos1959}, “watts-strogatz small-world” \cite{Watts1998} and “ scale-free networks”\cite{Albert2002}.\footnote{For a more comprehensive study of networks, look at \cite{Newman2010, Wasserman1994}}  All these networks have been visualized in Figure \ref{Random}, Figure \ref{SW} and Figure \ref{PA} respectively.\\
Using network structures as an ABM environment provides lots of opportunities to synthesize social network theory (SNT) with ABM. As a matter of fact, ABMs are developed to explain how simple rules generate complex emergence (i.e. a process model) and in terms of network science ABMs are used to analyze the pattern that arise from agents’ interactions over the time (i.e. a pattern model)\cite{wilensky2015introduction}. When spatial environment and networked environment come together, they form a mixed environment as visualized in Figure\ref{bblocks}.\\
As the third building block of ABMs, interactions refer to rules of behaviors for both agents and the environment\cite{Epstein1997}. Actually, these rules enable agents to interact with both themselves and others. There are five basic classes of interactions: agent-self, environment-self, agent-agent, environment-agent, and environment-environment. In agent-self interactions, an agent checks its internal states and decides according to them. Environment-self interactions are when areas of the environment alter or change themselves. For instance, they can change their internal state variables as a result of some calculations. Agent-agent Interactions are usually the most important type of action within ABMs. Agent-Environment Interactions happen when the agent manipulates or examines an area of the world in which it exists, or when the environment in some way observes or alter the agent’s internal states. Environment-Environment Interactions between different areas of the environment are probably the least commonly used interaction type in ABMs \cite{wilensky2015introduction}.
 \section{Development of ABMs} \label{section.development}
Generally, an ABM can be developed through three sequential and often iterative phases\footnote{There is not still a completely agreed-upon and standard protocol for ABM development among all practitioners\cite{Windrum2007}. One good protocol for ABM development is in\cite{VanDam2012}}:(1) designing phase, (2) programming phase and finally, (3) examination and analysis phase\cite{wilensky2015introduction}. In designing phase, the initial skeleton of ABMs is constructed. This initial skeleton is actually the textual model of ABM through which all behavioral rules and properties of agents and environment along with the way they interact with each other are verbally documented.  As the second phase, programming phase deals with how to translate the ABM’s textual model to a computational model by agent based programming languages and toolkits. The examination and analysis as the third phase are conducted for getting insights concerning (1) model verification, (2) model validation, (3) model replication and reproducibility and (4) model’s output analysis. In the following, these three phases would be discussed.
 \subsection{Designing Phase} \label{section.Development.Designing}
Through designing phase, all behavioral rules and properties of agents and environment along with the way they interact with each other are documented by natural language of subject matter experts (SMEs) in this phase. This documentation actually serves as a textual model.\footnote{A textual model is the conceptual model of an ABM which is documented in natural language.}8 basic stages have to be well addressed as detailed in Table\ref{table3}:
\begin{longtable}{| p{.25\textwidth} | p{.75\textwidth} |}
   \caption{Stages of ABM development}\label{table3}\\
   \hline
 Stage & Description  \\
 \hline
1-	Underlying questions of model & What questions does the model want to answer?
Which aspects of the real system under study are going to be described in the model?\\
   \hline
 2-	Types of agents and granularity & What types of agents are going to be created in the model? \\
   \hline
 3- Granularity of agents & What is the granularity of the agents?
Are agents coarse-grained or fine-grained or both of them?
\\
   \hline
 4- Properties of agents   & What properties do agents hold?\\
   \hline
      5-	Behavioral rules & What are the behavioral rules of agents?
How different are these behavioral rules?
\\
   \hline
      6-	Environment structure & What are the external forces affecting agents? Is environment spatial or networked or both of them?
\\
   \hline
      7-	Input parameters & What are input variables to model?
What is the type of input variables? (Boolean, string, continuous,..)
\\
   \hline
     8-	Outputs and measures & What measures and outputs are going to be collected from the model?\\
   \hline
     9-	scheduling & What is the sequence and time-step of model’s events?\\
   \hline
 \end{longtable}
While designing an ABM, three main factors associated with ABMs’ modeling approaches should be taken into consideration. Actually, In ABM literature, modeling approaches can be divided based on a number of aspects. Three of the most important aspects include 1-goal of modeling, 2-development method and 3-elaboration strategy. In terms of modeling goal, ABMs can be grouped into two major categories of phenomena-based modelling and exploratory modelling\cite{wilensky2015introduction}. In phenomena-based modeling researchers begin with a known target phenomenon. Typically, that phenomenon has a characteristic pattern, known as a reference pattern. Examples of reference patterns might be common housing segregation patterns in cities, diffusion of a specific ICT technology, spiral-shaped galaxies in space or oscillating population levels in interacting species\cite{wilensky2015introduction}. These reference patterns are those statistical regularities that econometricians suppose as stylized facts for example, the way price affects supply or demand. The goal of phenomena-based modeling is to create a model that will somehow capture the reference pattern. In ABM, this translates to finding a set of agents, and rules for those agents that will generate the known reference pattern. Once you have generated the reference pattern you have a candidate explanatory mechanism for that pattern and may also vary the model parameters to see if other patterns emerge, and perhaps try to find those patterns in data sets or by conducting experiments. Phenomena-based modeling can also be used with other forms of modeling, such as equation-based modeling. In equation-based modeling, this would mean writing down equations that will give rise to the reference pattern. Evidently all empirical validations perform well in case of phenomena-based modelling where there is a reference pattern against which the accuracy of model’s results is measured.\\
The second core modeling form is exploratory modeling. This form is perhaps less common in equational contexts than it is in ABM literature. In exploratory modeling with ABM, a researcher can create a set of agents, define their behavior, and explore the patterns that emerge. One might explore them solely as abstract forms, much like cellular automata developed by Conway\cite{Conway1976} but to count as a modeling practice, we must note similarities between the behavior of our model and some phenomena in the world just as patterns generated by cellular automata like oscillators and spaceships \cite{Wolfram1983}. Then our ABM should be refined in the direction of perceived similarities with these phenomena and converge toward an explanatory model of some phenomenon.\\
Phenomena-based modelling stems from the notion that there is an objective and real but unobservable data generating mechanism and that the purpose of any model is to represent elements of that mechanism in ways that generate some of the same data. But, in the case of exploratory based modeling the purpose of the models is the representation of perceptions of policy analysts and other stakeholders in the relevant social processes\cite{Moss2008}.  The phenomenon based modeling is a class of agent based models that has much in common with mainstream economic models. They incorporate utility functions; they employ numerical representations of phenomena and attributes naturally described in qualitative terms by the individuals being represented and by other stakeholders. The exploratory modeling is a class of models emerging from a process that is embedded in the social process of policy and strategy formation. Such models are typically described in linguistic terms (i.e. mental models) used by stakeholders rather than numerical variables intelligible and meaningful only to modelers. Such models are developed to facilitate stakeholder participation in the model design and validation process. They are intended precisely to represent the perceptions of stakeholders in order to bring clarity to scenarios built to explore the possibilities - the opportunities and threats - of an uncertain future. The major function of exploratory modeling is to enable the subject matter experts (SMEs) to see what outcomes their mental model(s) can produce when implemented in a real-world system. Therefore, they can have lots of advantages for those who deal with understanding complex systems.\\
In terms of elaboration strategy, ABMs can be grouped into KISS, KIDS and TAPAS strategies. KISS strategy that stands for “Keep It Simple, Stupid”. This notion is rooted in the Occam’s razor principle stating “while being faced with a number of competing hypotheses for a problem, one should select a hypothesis that has a fewer assumption” This principle advocates the law of parsimony and agent-based modelers that use this principle “start from simple models and gradually sophisticate it to answer their question”. KIDS strategy which stands for “Keep It Descriptive, Simple” is in opposite direction of KISS strategy. Advocates of KIDS strategy start from descriptive models and gradually simplify them to answer their questions\cite{Windrum2007,Elsenbroich2014}.\footnote{Technically, KISS and KIDS can be supposed as two opposite ends of spectrum. In an effort for developing a unified strategy, Rand and Wilensky (2007) developed full spectrum modeling strategy. In this strategy models are developed in a progressive way ( either from simple to descriptive or  descriptive to simple) and the phenomenon under study is modelled at multiple levels of details\cite{Rand2007}} TAPAS strategy that stands for (Take A Previous model and Add Something). In this strategy, modelers take an existing model and successively modify it through adding new features or relating initial assumption\cite{Frenken2006}.\\
In terms of development, ABMs can be grouped into two major categories of theory-based modelling and evidence-based modelling\cite{Moss2008}. ABMs can be developed via a theory. Actually a theory that specifies the behavioral rules of agents or the statistical regularities that the model is designed to explain them. Since theory-based ABMs are built upon prior empirical studies and aimed at stimulating real-world aggregates such as technology diffusion, disease spread and inflation formation, they are basically used for phenomena-based modeling where the modelling of a real-world pattern is the purpose of the modeler. Besides theories, ABMS can also be developed based on evidence.  The evidence-based modeling is used when researchers have a mental model concerning behavioral rules of agents of a system and they are interested in understanding the collective behavior of that system when its agents interact with each other. As it can be inferred, this approach is the foundation of exploratory modeling approach where a model is developed for representing the emergent properties of researcher’s mental models.\\
Evidence-based ABMs can be developed either as participatory modeling or individual thought experiments. One of major forms of participatory modeling is "companion modeling"(ComMod) which is an iterative participatory approach where multidisciplinary researchers and stakeholders work together continuously throughout a three-stage cycle \textit{field work $->$ modelling $->$ simulation $->$ field work} again\cite{Barreteau2003}.\\
ComMod follows two basic objectives. First, it increases the understanding of complex systems through its three-stage cycle. Second, it supports collective decision-making Processes in Complex Situations. In this case, the approach facilitates collective these kinds of processes by making more explicit the various points of view and subjective criteria, to which the different stakeholders refer implicitly or even unconsciously. Indeed, as demonstrated in past research\cite{Funtowicz1994}when facing a complex situation, the decision-making process is evolving, iterative, and continuous. It means that this process produces always imperfect "decision acts", but following each iteration they are less imperfect and more shared. Principally, the main principle of the ComMod approach is to develop simulation models integrating various stakeholders’ points of view and to use them within the context of platforms for collective learning. This is a modeling approach in which stakeholders participate fully in the construction of models to improve their relevance and increase their use for the collective assessment of scenarios. The general objective of ComMod is to facilitate dialogue, shared learning, and collective decision-making through interdisciplinary and “implicated” action-oriented research to strengthen the adaptive management capacity of local communities\cite{Barreteau2003}. As a software engineering method Virtual Overlay Multi Agent System (VOMAS) has been used to improve ComMod methodology\cite{niazi2012cognitive}.VOMAS is used for facilitating last two stages of ComMod, namely modeling and simulation. This method has been very successful in building verified and validated agent based models\cite{NiaziMuazA;HussainAmir;Kolberg2017}.In a nutshell, the textual model (i.e. the conceptual model) of an ABM is the output of designing phase.

 \subsection{Programming Phase} \label{section.Development.Programming}
When the textual model of an ABM is designed, it should be simulated through an agent based programming languages or simulation toolkits.\footnote{Sometime an ABM can be directly developed in the programming phase through participatory simulation platforms. This type of simulation is useful for simulating systems that there is not enough data about them therefore designing an initial conceptual model for them is almost impossible. According to this approach, an agent-based model is directly developed through direct participation of stakeholders of the problems in distributed platforms such as client-server network. This simulation is very useful in research and education\cite{Colella2000,Frey2013}.} Such simulation toolkits are a type of simulation software specifically for translating the textual model of an ABM into a computational model. A simulation is an understandable manifestation of a model, coded and visualized by a computer program which provides insights regarding the system under study. A simulation model basically refers to the computing algorithms or mathematical expression that entail the performance and total behavior of a system in the real world scenarios\cite{Abar2017}. In the early 1990s, general purpose programming languages (GPPLs) were basically used for simulation. SMALLTALK, C++ and Java were the most common GPPLs in ABM community of practice. Using GPPLs for agent based simulation have some visible disadvantages. For example, modelers have to implement basic functions and plotting from scratch and they should be very familiar with the programming language\cite{Gilbert2002}. Such problems have resulted in development of agent-based simulation toolkits which help modelers a lot to simulate the complex system under study. As presented in Table\ref{Toolkits and Languages}, the majority of toolkits support the primary GPPLs including Java, C++, C and Logo variant.
\begin{longtable}{|c|c|c|c|c|}
 \caption{Toolkit and their programming languages adapted from\cite{Nikolai2009}}
\label{Toolkits and Languages}\\
    \hline
  \multirow{2}{0.5cm}{Toolkit} & \multicolumn {3}{c} {Programming Language} &  \\
  \cline{2-5}
  & Java &C++ & C & Logo\\
    \hline
  ADK &  * & &  & \\
   \hline
   AgentBuilder & * &*  &* & \\
    \hline
    AnyLogic & *& &  & \\
    \hline
     Ascape &* &  & & \\
    \hline
     DeX & & * & & \\
    \hline
    Echo & &  &* & \\
    \hline
    iGen & *& * &* & \\
    \hline
     LSD & & * & & \\
    \hline
     MadKit &* & *& *& \\
    \hline
    MAML & &  & *& \\
    \hline
    Mason &* &  & & \\
    \hline
    Netlogo & &  & &* \\
    \hline
    RepastS &* &  & & \\
    \hline
  Starlogo & &  & &* \\
    \hline
  StarlogoT & &  & &* \\
    \hline
  Swarm &* &  & & \\
    \hline
 \end{longtable}
As it can be seen in above table, some of platforms are supported by multiple languages such as AgentBuilder, iGen and Madkits that are supported by Java, C++ and C. During this paper, Nelogo 6.0.1 has been used to simulate all ABMs. As one of the most frequently used agent based modeling and simulation toolkit, Netlogo was developed by Uri Wilensky in 1999. Since its development, it has been regularly updated in sequence of versions and a number of extensions.\footnote{Readers can refer to Netlogo home page \uline{ccl.northwestern.edu/netlogo/} in order to get more information of this agent based modeling toolkit.}  When model designers and model programmers are different persons, before offering their conceptual models to programmers, model authors had better put it in a unified modeling language (UML) format. Because, it helps programmers a lot to accurately discern what model authors want\cite{Bersini2012}.
 \subsection{Examination Phase} \label{section.Development.Examination}
 When ABMs are designed and programmed, they should be examined. Examination phase is highly critical for showing (1) is ABM right designed? (2) Is a right ABM designed? (3) Is ABM replicable by other researchers? And (4) how should outputs be analyzed? Each of these questions will be addressed in the following:
 \subsubsection{Verification } \label{section.Development.Examination.Verification}
 In verification process the main purpose is to determine whether the designed model corresponds to programmed model. As it is sensible, through verification process the researchers try to understand the gap between the designed model and its implementation (computational model) and fill it through correction and code debugging. As a matter of facts, verification process comes to play when the model author (i.e., subject matter expert or designer) and model implementer (i.e., simulation specialist or programmer) are different persons which is very common among academic researchers cooperating in a team\cite{NiaziMuazA;HussainAmir;Kolberg2017,niazi2012cognitive,niazi2017towards,Rand2011,wilensky2015introduction}. In such a situation, the model author should iteratively discuss the model with the model programmer so that the model programmer understands and programs what exactly the model author wants. This process is called “iterative modelling”\cite{wilensky2015introduction}which can help model verification process when model author and model programmer are different persons (See the backward curve with the red arrow in Figure\ref{ABMDevelopmentProcess}). One very noteworthy point for facilitating the verification process is that the SMEs frequently check textual model developed in designing phase and find its inconsistencies with the computational model developed in programming.
 \subsubsection{Validation } \label{section.Development.Examination.Validation}
 Validation of computational models has always been a major concern for Simulation specialists\cite{Carley1996,Garcia2007,Rand2011,Windrum2007}. Validation is defined as when the simulated model produces the results that are in a satisfactory range of accuracy matching up with the real-world data\cite{Windrum2007}. Model validation is the process of determining whether the implemented model corresponds to, and explains some phenomenon in the real world\cite{wilensky2015introduction}. When a model is implemented, its validation becomes so essential. A valid model assures the researchers of the model’s rightness\cite{PullumLauraL;Cui2012}. Therefore, the model’s results are supposed to be useful out of the model and can be confidently used for policy making.  A number of studies have been conducted about types of validation\cite{PullumLauraL;Cui2012,Windrum2007}, levels through which validation can be done\cite{Rand2011,wilensky2015introduction}and various methodologies for conducting validation\cite{Moss2008,NiaziMuazA;HussainAmir;Kolberg2017,Windrum2007}. In a big picture, all validation methods can be classified as qualitative and quantitative methods. qualitative methods like face validity are very subjective and expert-based while quantitative ones such as empirical validation approaches are objective and based on real-world data\cite{PullumLauraL;Cui2012} (Pullum, Laura L; Cui, 2012). Face validation is the process of showing that the mechanisms and properties of the model look like mechanisms and properties of the real world. It is mainly conducted by subject matter experts (SMEs). Empirical validation makes sure that the model generates data that can be demonstrated to correspond to similar patterns of data in the real world. Validation approaches of ABMs essentially depend on modelling approaches\cite{Moss2008}. When ABMs are used to simulate a real-life statistical regularity as is the case of phenomena-based modeling, empirical validation approaches are used. But when ABMs are used to simulate the mental models of systems stakeholders and show what will emerge from them as is the case of exploratory modeling, qualitative validation methods such as face validity are used. Since ABMs have different micro-macro behavior levels (e.g., an agent’s individual behavior vs model’s behavioral aggregate),they should be validated according to two axes\cite{Rand2011}as shown in Table\ref{Validation Axes}.
\begin{center} 
\begin{table}
\centering
\caption{Validation Axes}
\label{Validation Axes}
  \begin{tabular}{|c|c|c|c|}
\hline

 \multirow{4}{1.2cm }{Axis 2} & Macro&  Macro face Validation& Macro empirical Validation \\[8.5pt]

\cline{2-4}
  &Micro&Micro face Validation& Micro empirical Validation\\

\cline{2-4}
  &  & Face& Eimpirical\\

\cline{2-4}
  \multicolumn{4}{|c|}{Axis 1}\\
  \hline
  \end{tabular}
  \end{table}
\end{center}
 In the macro –face validation, the SMEs want to know how much the aggregated patterns produced by the implemented model correspond “on face” to the real-world aggregated patterns? In the micro-face validation, the SMEs plan to discern how much properties and behavior rules of agents correspond “on face” to reality? In the macro-empirical validation, the SMEs want to know how much the outputs or targeted aggregates produced by the implemented model correspond to real-word data? In the micro-empirical validation, the SMEs want to discern how much input parameters, properties and behavior rules of agents correspond to real-word data? In all empirical validations, most often the model should be calibrated both at micro and macro levels. The calibration is a process through which the suitable parameters and initial conditions are found so that the implemented model generates a pattern similar to real-world pattern\cite{Carley1996,Rand2011,Windrum2007,wilensky2015introduction}.
 \subsubsection{Replication } \label{section.Development.Examination.Replication}
 If a model is going to be acceptable among a scientific community, it must be replicable. Replication of a model is actually the re-implementation of its conceptual model according to the previous results produced by its implemented model. When a computational model is replicated, it not only shows its verification is trustable but also reexamines its reevaluation and  facilitates a common language among modelers\cite{wilensky2015introduction}. One effective way for facilitating the replication of a model is publication of its programming codes by its authors in open access platforms such as\uline{ http://modelingcommons.org/account/login} for Netlogo programmers and\uline{ https://www.comses.net/ }for all programmers. Another effective way is a full publication of (1) hardware specifications of a computer used for simulation and (2) the programming language and toolkits used for simulation. However, one very noticeable point is the fact that when a model is successfully replicated it should be able to produce outputs A successful replication is one in which the replicators are able to establish that the replicated model creates outputs sufficiently similar to the outputs of the original. This does not necessarily mean that the two models have to generate the exact same results\cite{wilensky2015introduction}.\footnote{For a more comprehensive study of  ABMs’ replication , look at\cite{Axtell1996}.}
 \subsubsection{Output Analysis} \label{section.Development.Examination.Analysis}
\begin{center}
\begin{figure}
\centering
    \includegraphics[scale = 0.55,trim = {0cm 0cm 2cm 0cm}, clip = true]{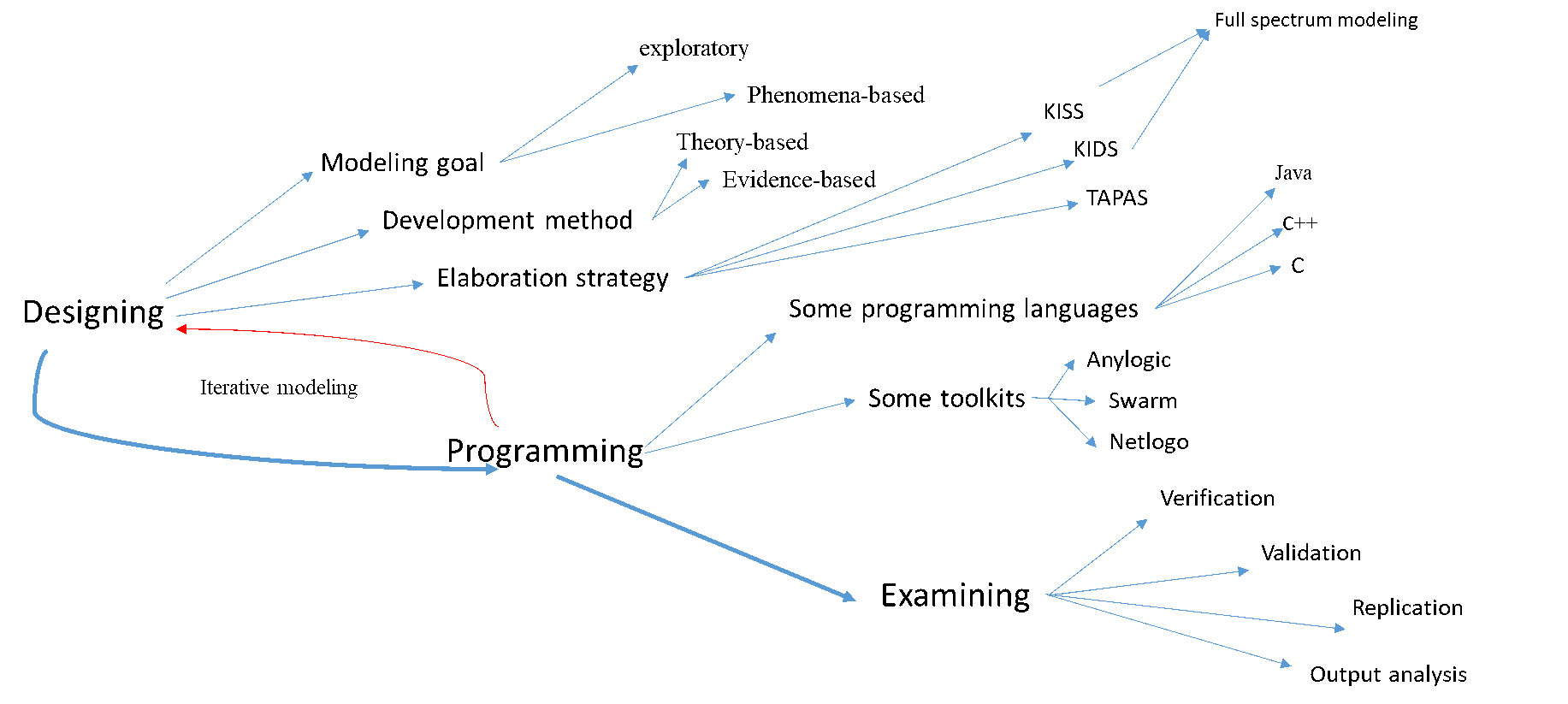}
    \caption{ABM Development Process}
    \label{ABMDevelopmentProcess}
\end{figure}
\end{center}

ABMs generate a great deal of data both at micro-level and macro-level. Because of such a data comprehensiveness, a number of analyses can be conducted on such models. As the eighth stage of table 3, the output analysis is related to measures that the SMEs want to analyze the effect of model input parameters on them. One very noteworthy point in analysis of ABMs’ outputs is the stochastisity of such computational models. Actually, like CASs, ABMs have stochastic nature. Even with the same set of parameters (parameter combinations), They produce different results in different runs. This requires researchers to do such simulations in multiple runs and analyze statistical distributions of targeted results (behavior under study) through inferential statistics methods\cite{axtell2000agents,wilensky2015introduction}. To have a good analysis, in addition to descriptive statistics measures, the researchers can use inferential statistics methods such as various statistical tests in analyzing the statistical distribution of ABMs’ outputs. Moreover, if a network environment is used, several number of social network theory (SNT) measures can be very useful for extracting specific insights from ABMs’ outputs. In the nutshell, the big picture of ABM development process is visualized in Figure\ref{ABMDevelopmentProcess}.
\section{Two Critical Considerations} \label{section.considerations}
\begin{center}
\begin{figure}
\centering
    \includegraphics[scale = 0.6,trim = {0cm 0cm 0cm 0cm}, clip = true]{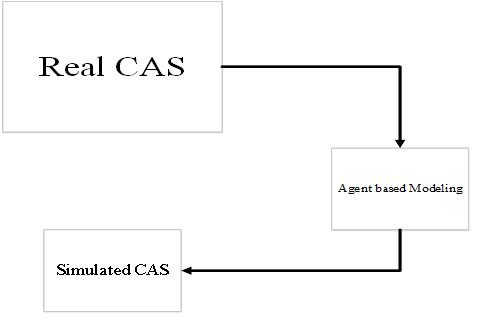}
    \caption{A real CAS and a simulated CAS}
    \label{realsimulated}
\end{figure}
\end{center}

Complex adaptive systems (CASs) are real world systems which have a number of characteristics (As discussed in Table\ref{table1}. In contrast, Agent based models (ABMs) are a type of computational methodology believed to be very promising in modeling CASs. Every developed ABM only shows one or some aspects of a CAS not all of its aspects. Therefore, what ABM practitioners produce via simulating a CAS (e.g., a society) is a simplified and artificial picture of that CAS as it is visualized in Figure\ref{realsimulated}. Two significant factors that have to be taken in to consideration by every ABM practitioner would be discussed in the following.

 \subsection{Emergence} \label{section.considerations.Emergence}
 CASs exhibit emergent properties. A property of a CAS that emerges out of non-linear (non-trivial) interactions among its constituent components so that it is beyond and irreducible to them is called “emergence”.  In comparison to other simulation techniques such as discrete event simulation (DES), system dynamics (SD) or even game theory, one of the greatest advantages of ABM is its outstanding prowess in showing the emergent properties of CASs. Our world exhibits several observable CASs, in biological sciences, molecules emerge out of interacting atoms, organelles emerge out of interacting molecules, cells emerge out of interacting organelles. Organs emerge of interacting cells and finally body emerges out of interacting organs. Such examples of emergent properties of our body indicates that it is a biological CAS full of interesting emergent properties. In a sociological perspective, there are also several examples of CAS, as an instance, the society can be interpreted as an emergent property of interacting humans or even norms can be studied as emergent property of social system. The emergence is very difficult to forecast and completely depends on the observation\cite{sun2006cognition}. In facing the challenge of emergence, there can be two kinds of human thinking way\cite{wilensky2015introduction}.

\begin{figure} 
    \centering
    \begin{subfigure}{0.7\textwidth}
        \includegraphics[scale = 0.55,trim = {0cm 0cm 0cm 0cm}, clip = true]{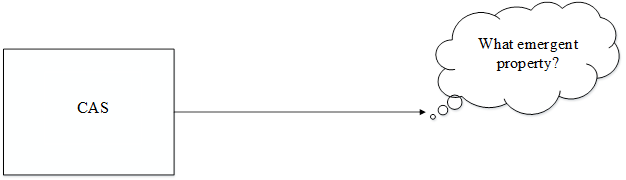}
        \caption{Integrative Thinking}
        \label{Integrative}
    \end{subfigure}
 ~
    \begin{subfigure}{0.7\textwidth}
        \includegraphics[scale = 0.55,trim = {0cm 0cm 0cm 0cm}, clip = true]{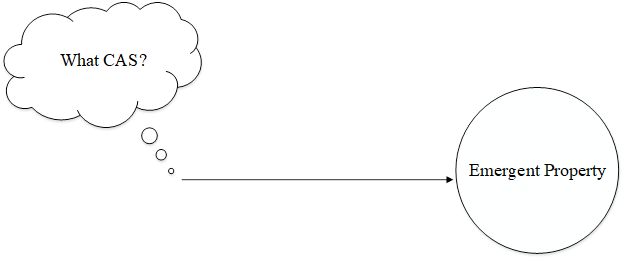}
        \caption{differential Thinking}
        \label{differential}
    \end{subfigure}
    ~
    \caption{Two different kinds of Thinking}\label{thinking styles}
    \end{figure}     

As shown in \ref{Integrative}, integrative thinking refers to when the SMEs have a CAS in mind (e.g., an organization into which a number of people with different religious backgrounds and specialties work) but they don’t know its targeted emergent property (e.g., the pattern of cooperation or formation of hierarchy). In contrast, when SMEs have an observable emergent property but don’t know its CAS (i.e., from which CAS, that property emerged) they are facing differential thinking as visualized in \ref{differential}.\\
 Essentially, integrative thinking comes to play when the exploratory modeling is assumed while differential thinking takes effect when the phenomena-based modeling is adopted. However, whereas the core of integrative thinking is to discern what properties will emerge out of the CAS under study, differential thinking is used to grasp what CASs can lead to the emergent property under study.

 \subsection{Non-Ergodicity of ABMs and the necessity of sensitivity analysis} \label{section.Development.Non-ergodicity}
 Like CASs, ABMs are non-ergodic. Non-ergodic systems are highly sensitive to initial conditions. If their initial conditions are a little changed, their output behaviors will be drastically influenced. So, it is necessary to do sensitivity analysis in ABMs after they are programmed and evaluate how their final behaviors (measures) are sensitive to change of their input parameters\cite{Al-suwailem2008, wilensky2015introduction }. A good sensitivity analysis can help SMEs (1) to get aware of how emergent properties are produced in ABMs, (2) to get able to examine the robustness of emergent properties and (3) to get able to quantify the changeability of ABM’s outputs after input parameters are changed. A number of methods have been proposed for sensitivity analysis of ABMs, three important if which are (1) one-factor-at-a-time (OFAT), model-free output decomposition of variance and (3) variance decomposition of model based output\cite{TenBroeke2016}.
 \section{Two Simulations} \label{section.simulations}
 To illustrate the implementation of ABM, two economic systems have been simulated using an exploratory agent-based modeling approach\cite{Sabzian2018}. Exploratory ABM has a wide potentiality for thought experiment\cite{Rangoni2014}. As discussed above, this approach is grounded on evidence based modelling where a SME (or a team of them) likes to make sense of what will possibly emerge out of their mental models before being executed in real world. Outcomes produced by an exploratory ABM can find a number of empirical supports in real world data. Therefore, it can also play a vital role in theory development\cite{Macy2002}.Through this implementation, an exploratory ABM has been used to show (1) how economic inequality emerges within an economic system (2) how charity and allocation strategies of charity entities can help reduce this emergent economic inequality.\\
The firs system (system I) is mainly inspired by the work of Wilensky and Rand (2015). However, to make this model more suitable for this study, some new features have been added to its original version including (1) the possibility of changing the number of agents in the interface view, (2) the possibility of choosing money amount for agents in the interface view, (3) the possibility of simulation when agents start with equal or unequal amount of money in initial condition (see \ref{equal amount} for equal amount and see \ref{unequal amount} for unequal amount) and (4) the possibility of showing the amount of money of each agent by its color in the world view so that the richer agents get a darker color and go north wise while the poorer agents get a lighter color and go south wise (Figure \ref{unequal amount}).\\
The second system (system II) is an extension of system I which includes a number of new features such as (1) the ability of human agents to give charity, (2) the establishment of charity agents, (3) the strategies that charity agents (entities) can take to allocate the charity among the needy agents and so on . In order to make the replication of this model easier, the agent-based simulation of system II can be directly searched and downloaded from{ \scriptsize{\verb"http//modelingcommons.org/browse/one_model/5379#model_tabs_browse_info"}} that is a useful platform for communicating and discussing agent-based models written in Netlogo 6.0.1. The conceptual models (textual models) of system I and system II along with their simulation and analysis would be discussed in the following:
\begin{figure}
    \centering
    \begin{subfigure}{0.7\textwidth}
        \includegraphics[scale = 0.55,trim = {8cm 0cm 8cm 4.5cm}, clip = true]{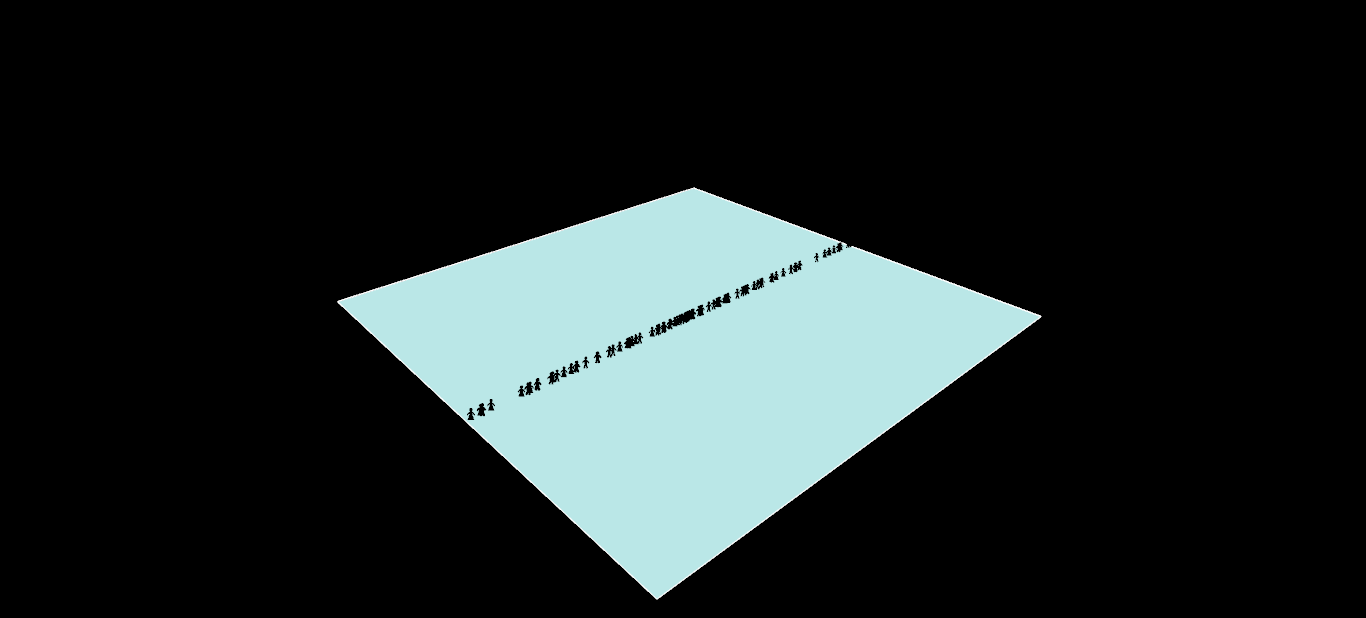}
        \caption{A 3D view of 100 agents with equal amount of money in the initial conditions}
        \label{equal amount}
    \end{subfigure}
 ~
    \begin{subfigure}{0.7\textwidth}
        \includegraphics[scale = 0.55,trim = {8cm 0cm 8cm 4.5cm}, clip = true]{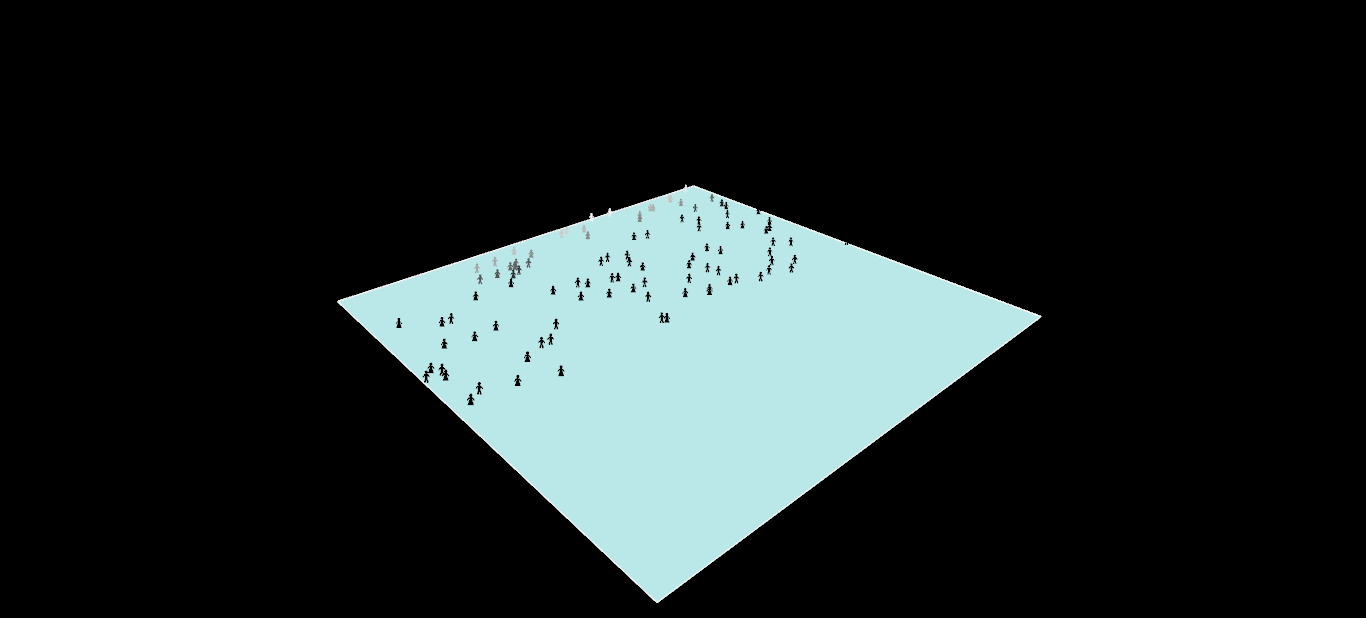}
        \caption{A 3D view of 100 agents with unequal amount of money in the initial conditions}
        \label{unequal amount}
    \end{subfigure}
    ~
    \caption{Agents with equal and unequal amount of money in the initial conditions }\label{equal and unequal}
\end{figure}
\subsection{System I}\label{section.simulations.sysI}
\begin{figure}
    \centering
    \begin{subfigure}{0.7\textwidth}
        \includegraphics[scale = 1,trim = {0cm 0cm 0cm 0cm}, clip = true]{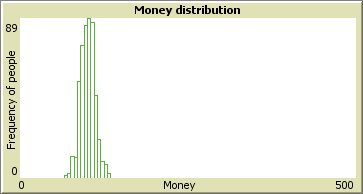}
        \caption{Probability distribution of money in tick 100}
        \label{100 ticks}
    \end{subfigure}
 ~
    \begin{subfigure}{0.7\textwidth}
        \includegraphics[scale = 1,trim = {0cm 0cm 0cm 0cm}, clip = true]{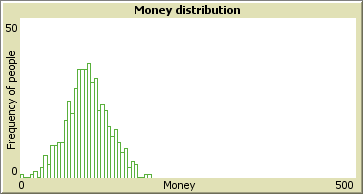}
        \caption{Probability distribution of money in tick 1000}
        \label{1000 ticks}
    \end{subfigure}
 ~
     \begin{subfigure}{0.7\textwidth}
        \includegraphics[scale = 1,trim = {0cm 0cm 0cm 0cm}, clip = true]{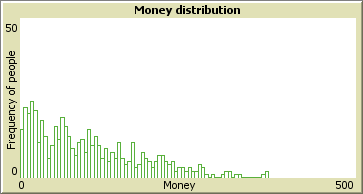}
        \caption{Probability distribution of money in tick 9000}
        \label{9000 ticks}
    \end{subfigure}
 ~
    \caption{Distribution of money in different ticks (100,1000,9000)}\label{Different ticks}
\end{figure}

\begin{figure}
    \centering
    \begin{subfigure}{0.7\textwidth}
        \includegraphics[scale = 1,trim = {0cm 0cm 0cm 0cm}, clip = true]{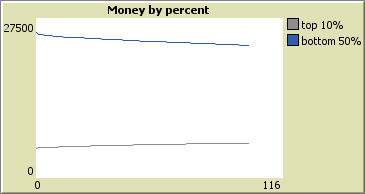}
        \caption{Distance of bottom 50\% from top 10\% in tick 100}
        \label{Distance in 100 ticks}
    \end{subfigure}
 ~
    \begin{subfigure}{0.7\textwidth}
        \includegraphics[scale = 1,trim = {0cm 0cm 0cm 0cm}, clip = true]{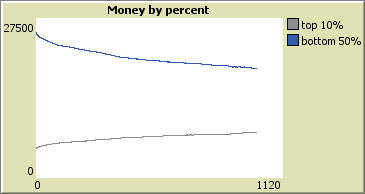}
        \caption{Distance of bottom 50\% from top 10\% in tick 1000}
        \label{Distance in 1000 ticks}
    \end{subfigure}
 ~
     \begin{subfigure}{0.7\textwidth}
        \includegraphics[scale = 1,trim = {0cm 0cm 0cm 0cm}, clip = true]{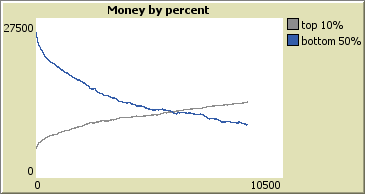}
        \caption{Distance of bottom 50\% from top 10\% in tick 9000}
        \label{Distance in 9000 ticks}
    \end{subfigure}
 ~
    \caption{Distance of bottom 50\% from top 10\% in different ticks (100,1000,9000)}\label{Distance}
\end{figure}

 This system is built upon five assumptions including (1) there is a society in which N number of persons live, (2) there is a global clock that shows the time by each tick and it is completely discrete, (3) each person has M amount of money that can be equal or unequal in the initial conditions, (4) there is a money gap between the five low deciles (bottom 50\% and the tenth decile (top 10\%). It can be positive, zero or negative. When it is positive, it shows the money of five low deciles is more than that of tenth decile. In case of being zero, it shows the money of five low deciles is equal to that of tenth decile and if it is negative, it indicates the money of five low deciles is less than that of tenth decile, (5) there is a critical threshold which shows the criticality of economic situation when the value of money gap becomes lower than it. According to this system, if all people (N = 500) of the society have an equal amount of money (M =100) in the initial conditions and donate a unit of money to each other randomly and by each time tick as long as each person’s money is more than zero, what would be the answers to following questions:
1-	How will the probability distribution of money be in tick of 100?
2-	How will the probability distribution of money be in tick of 1000?
3-	How will the probability distribution of money be in tick of 9000?
Because of the stochastic nature of agent-based models, the probability distribution of money in system I has been simulated in 10 different runs . In run 1, According to \ref{100 ticks}, After 100 ticks, the probability distribution of money is similar to a normal distribution as M ~ N (100, 110.612). This distribution shows that money of each person is very inclined to the average money of society. Therefore, all deciles of society have a somewhat similar amount of money. In addition, as shown in \ref{Distance in 100 ticks},there is a great distance between the money volume of all five lower deciles (bottom 50\% with 22885 units of money) and that of tenth decile (top 10\% with 5933 units of money) meaning that bottom 50\% has 16952 units of money more than top 10\%.\\
As demonstrated in r\ref{1000 ticks}, the probability distribution of money has become flatter when the system I is in tick of 1000. This normal distribution has a mean of 100 and variance of 972.661 implying that the majority of the society have a money inclined to 100 units of money and just a few of them own a very high amount of money. However, regarding \ref{Distance in 1000 ticks} it can be concluded, the total amount of money of bottom 50\% has become mitigated to 18835 while the top 10\% has accumulated a remarkable amount of money 7783 and money distance has decreased by a large degree to 11052.
According to \ref{9000 ticks}, when system I comes to tick of 9000, it shows a Boltzman-Gibbs distribution implying that just a very few of persons have accumulated a great amount of money and a large number of them have just gained a little amount of money. In this case the bottom 50\% has an amount of money of 9134 while top 10\% has an amount equal to 13052 that is 3918 units more than that of bottom 50\%. The money volume of top 10\% has exceeded that of bottom 50\% in tick of 5800 since then the gap has increased (see \ref{Distance in 9000 ticks}).
Statistical summary of all 10 runs for ticks of 100, 1000 and 9000 is described in Table\ref{System up to tick of 100}, Table\ref{System up to tick of 1000} and Table\ref{System up to tick of 9000} respectively.
  \begin{longtable}{|p{.1\textwidth} | p{.1\textwidth} |  p{.1\textwidth} |p{.1\textwidth} |p{.1\textwidth}|p{.1\textwidth}| p{.1\textwidth}|}
   \caption{System up to tick of 100}\label{System up to tick of 100}\\
 \hline
   \# of runs & Mean &Variance &Total money of top 10\% &Total money of bottom 50\% &diff &Tick of critical stage\\
     \hline
RUN 1 &100&110.621&5933&22885&16952&-\\
   \hline
RUN 2 &100&104.561&5877&22959&17082&-\\
   \hline
RUN 3 &100&102.997&5908&22988&17080&-\\
   \hline
RUN 4 &100&92.641&5824&23114&17290&-\\
   \hline
RUN 5 &100&100.917&5867&22974&17107&-\\
   \hline
RUN 6&100&97.002&5850&23025&17175&-\\
   \hline
RUN 7 &100&89.254&5840&23132&17292&-\\
   \hline
RUN 8 &100&90.509&5884&23149&17265&-\\
   \hline
 RUN 9 &100&100.044&5913&23028&17115&-\\
   \hline
  RUN 10 &100&106.052&5963&22987&17024&-\\
   \hline
 \end{longtable} 
\begin{longtable}{|p{.1\textwidth} | p{.1\textwidth} |  p{.1\textwidth} |p{.1\textwidth} |p{.1\textwidth}|p{.1\textwidth}| p{.1\textwidth}|}
   \caption{System up to tick of 1000}\label{System up to tick of 1000}\\
   \hline
   \# of runs & Mean &Variance &Total money of top 10\% &Total money of bottom 50\% &diff &Tick of critical stage\\
    \hline
RUN 1 &100&972.661&7783&18835&11052&-\\
   \hline
RUN 2 &100&950.793&7578&18871&11293&-\\
   \hline
RUN 3 &100&1031.490&7926&18555&10629&-\\
   \hline
RUN 4 &100&1092.669&7763&18279&10516&-\\
   \hline
RUN 5 &100&1054.322&7837&18425&10588&-\\
   \hline
RUN 6&100&986.492&7645&18778&11133&-\\
   \hline
RUN 7 &100&1001.531&7975&18765&10790&-\\
   \hline
RUN 8 &100&1259.535&8191&17949&9758&-\\
   \hline
 RUN 9 &100&958.436&7712&18822&11110&-\\
   \hline
  RUN 10 &100&1063.314&7840&e18480&0640&-\\
   \hline
 \end{longtable} 
\begin{longtable}{|p{.1\textwidth} | p{.1\textwidth} |  p{.1\textwidth} |p{.1\textwidth} |p{.1\textwidth}|p{.1\textwidth}| p{.1\textwidth}|} 
   \caption{System up to tick of 9000}\label{System up to tick of 9000}\\
   \hline
\# of runs & Mean &Variance &Total money of top 10\% &Total money of bottom 50\% &diff &Tick of critical stage\\
 \hline
RUN 1 &100&6256.793&13052&9134&-3918&5800\\
   \hline
RUN 2 &100&5359.6312&12692&10807&-1885&6476\\
   \hline
RUN 3 &100&5042.136&12207&10878&-1329&6310\\
   \hline
RUN 4 &100&4959.899&11890&10662&-1228&6919\\
   \hline
RUN 5 &100&5443.066&12562&10225&-2337&5801\\
   \hline
RUN 6&100&4869.382&11871&10737&-1134&6941\\
   \hline
RUN 7 &100&5597.174&12459&9723&-2736&5520\\
   \hline
RUN 8 &100&5326.837&12560&10373&-2187&5752\\
   \hline
 RUN 9 &100&5533.523&12648&10046&-2602&2725\\
   \hline
  RUN 10 &100&5469.022&12428&10011&-2417&5668\\
   \hline
 \end{longtable} 
As a fundamental law of equilibrium statistical mechanics, Boltzman-Gibbs law states that any conserved quantity in a big system should follow an exponential probability distribution. According to Boltzman-Gibbs law, in a closed economic system, the total amount of money is conserved because it is not manufactured, consumed or destroyed.
\begin{equation}\label{1}
p(m) = {{Ce} ^ \frac{-m}{t}}
\end{equation}
Here m stands for money, C is a normalizing constant and T is an effective temperature equal to the average amount of money per agent. Some studies have shown that the money distribution follows a Boltzman-Gibbs law when it is conserved and exchanged in a closed system. According to this property of money distribution a very few of persons will accumulate a great amount of money while a large number of them just gain a little amount of money\cite{Dragulescu2000, Ferrero2004,Yakovenko2009}. As it can be seen in table 3, in all ten runs, the top 10\% has outweighed the bottom 50\% in terms of accumulated money and consequently economic situation entered a critical stage in all runs. Thus, in order to prevent this economic system from becoming more critical, system I has been extended into system II by adding some more mechanisms to it.
 \subsection{System II}\label{section.simulations.sysII}
\begin{center}
\begin{figure}
\centering
    \includegraphics[scale = 0.55,trim = {6cm 0.5cm 9cm 5cm}, clip = true]{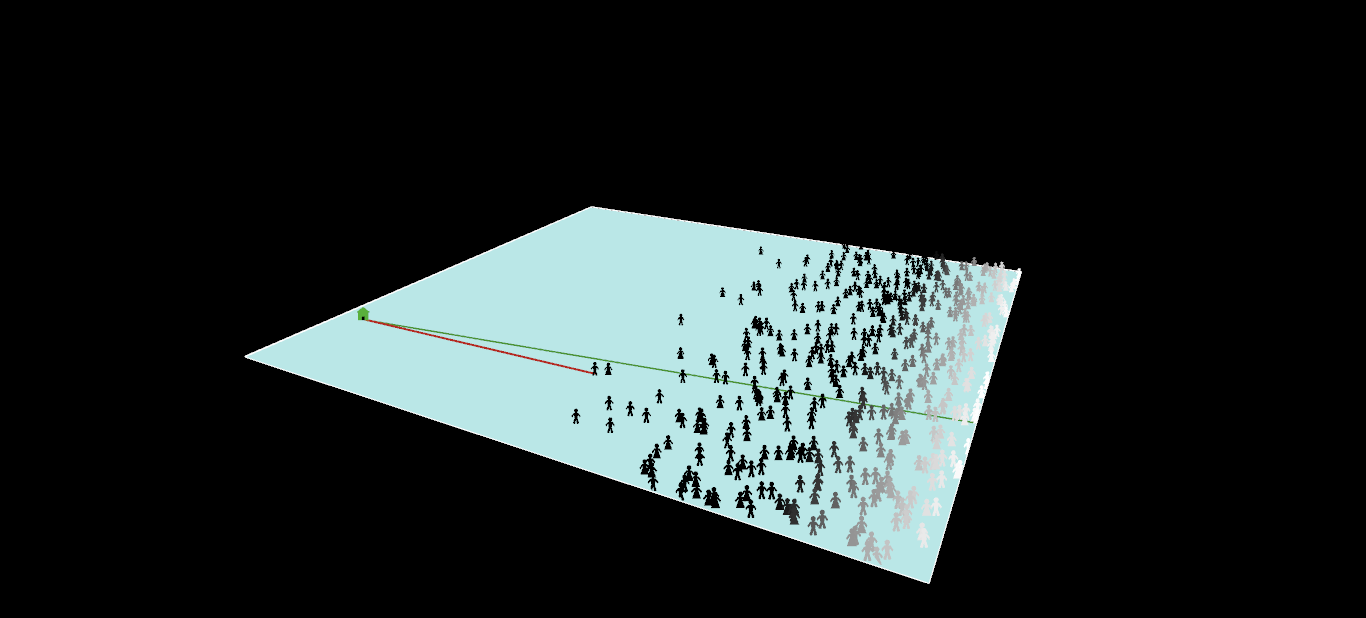}
    \caption{A 3D view of distribution of charity by charity entity while using strategy A in tick 5800}
    \label{Strategy A in tick 5800}
\end{figure}
\end{center}

 M number of charity organizations are added to system I. These organizations come to scene when the economic situation becomes critical (i.e., the money of 10\% of the society becomes equal to or more than that of its 50\%). The mission of these organizations is to help the low five deciles not to deteriorate more in the depth of poverty. These beneficiaries work based on the charity amount that benefactors give to them in order to distribute it among five lower deciles. In terms of charity -giving and distribution, these charities can take one of three allocation strategies of A, B or C. Strategy A is applied when just the richest person of the society gives a unit of his or her money to the charity organization and it allocates that money to the poorest person of the society.  When c\% of members of decile 10 give a unit of money to the charity organization and it allocates that of amount of money among d\% of members of five lower deciles, the charity organization has used strategy B. The charity entities use strategy C when k\% of members of decile 10, p\% of members of decile 9 and v\% of members of decile 8 give money (everybody one unit of money) to them and they distribute that amount of money among x\% of members of decile 1, y\% of members of decile 2 and z\% of members of decile 3 respectively. This model can help answering the following questions:
1-	How will strategy A affect the economic system when it enters a critical stage?
2-	How will strategy B affect the economic system when it enters a critical stage?
3-	How will strategy C affect the economic system when it enters a critical stage?
According to table 3, for run 1, the system has entered the critical stage in tick of 5800. The charity organization has used strategy A to help system exit this stage but it has not got out of the critical stage in the next ticks (up to 9000). It means that as long as the charity entity uses strategy A to help system get out of the critical stages, it fails to exit and returns to critical stage by each tick. The return period or recurrence interval of the critical stage is a key indicator for measuring the sustainability of allocation strategies. The visualization of how charity organization uses strategy A in tick of 5800 is shown in Figure\ref{Strategy A in tick 5800}.
According to Figure\ref{Strategy A in tick 5800}, the red line indicates charity from the richest person of the society (benefactor) to charity and the green line indicates the charity distributed by charity to the poorest person of the society. As it can be noticed, the spatial position and color of rich persons are different from those of the poorer ones. Such differences are because of a mechanism embedded into this model which forces the rich get darker color and move north wise while making the poor get brighter and move south wise. As shown in figures 20 to 25, for run 1, strategy A, strategy B (with parameters of c=100 and d= 20) and Strategy C (with parameters of k = 100, p = 60, v= 40, x= 100, y=60 and z= 40) have been applied for handling the economic critical stages up to 9000 ticks.
 \begin{figure}
    \centering
    \begin{subfigure}{0.7\textwidth}
        \includegraphics[scale = 1,trim = {0cm 0cm 0cm 0cm}, clip = true]{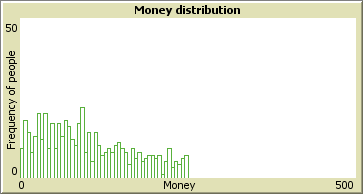}
        \caption{Distribution of charity by charity entity while using strategy A for 9000 ticks}
        \label{Distribution and Strategy A for 9000 ticks}
    \end{subfigure}
 ~
    \begin{subfigure}{0.7\textwidth}
        \includegraphics[scale = 1,trim = {0cm 0cm 0cm 0cm}, clip = true]{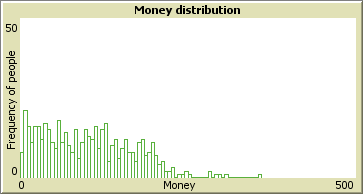}
        \caption{Distribution of charity by charity entity while using strategy B for 9000 ticks}
        \label{Distribution and Strategy B for 9000 ticks}
    \end{subfigure}
 ~
     \begin{subfigure}{0.7\textwidth}
        \includegraphics[scale = 1,trim = {0cm 0cm 0cm 0cm}, clip = true]{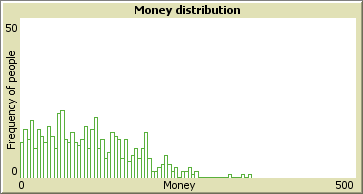}
        \caption{Distribution of charity by charity entity while using strategy C for 9000 ticks}
        \label{Distribution and Strategy C for 9000 ticks}
    \end{subfigure}
 ~
    \caption{Influence of different strategies(A,B,C) in charity distribution}\label{Distribution and Strategy}
\end{figure}
As demonstrated in \ref{Distribution and Strategy A for 9000 ticks},\ref{Distribution and Strategy B for 9000 ticks} and \ref{Distribution and Strategy C for 9000 ticks}, by each of allocation strategies, charity organization have tried to help the economic system exit the critical stage while entering it in every time. In this case when the total money of top 10\% becomes equal to or exceeds that of bottom 50\%, the total money of top 10\% will be forced back to a value less than that of bottom 50\% depending on how many of three higher deciles participate in giving charity and also which allocation strategies the charity organizations use to distribute the charity among poorer deciles. According to the Boltzman-Gibbs law after one of allocation strategies is implemented the money (as a conserved quantity here) will tend to become accumulated into hands of a very few of people. Thus the systems enter the critical stage again. The number of times that economic system returns to the critical stage after a typical strategy is used for resource redistribution in it, is a key factor for measuring how sustainable that strategy is.
 \begin{figure}
    \centering
    \begin{subfigure}{0.7\textwidth}
        \includegraphics[scale = 1,trim = {0cm 0cm 0cm 0cm}, clip = true]{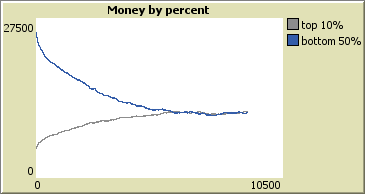}
        \caption{Distance of bottom 50\% from top 10\% in tick 9000 when the charity entity uses strategy A}
        \label{Distance and Strategy A for 9000 ticks}
    \end{subfigure}
 ~
    \begin{subfigure}{0.7\textwidth}
        \includegraphics[scale = 1,trim = {0cm 0cm 0cm 0cm}, clip = true]{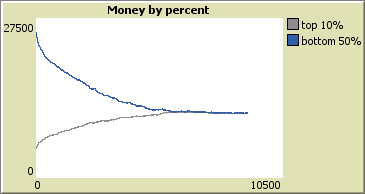}
        \caption{Distance of bottom 50\% from top 10\% in tick 9000 when the charity entity uses strategy B}
        \label{Distance and Strategy B for 9000 ticks}
    \end{subfigure}
 ~
     \begin{subfigure}{0.7\textwidth}
        \includegraphics[scale = 1,trim = {0cm 0cm 0cm 0cm}, clip = true]{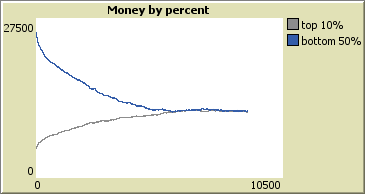}
        \caption{Distance of bottom 50\% from top 10\% in tick 9000 when the charity entity uses strategy C}
        \label{Distance and Strategy C for 9000 ticks}
    \end{subfigure}
 ~
    \caption{Distance of bottom 50\% from top 10\% after executing each strategy (A,B,C)}\label{Distance and Strategy}
\end{figure}
As it can be seen from \ref{Distance and Strategy A for 9000 ticks},\ref{Distance and Strategy B for 9000 ticks},\ref{Distance and Strategy C for 9000 ticks}, when a typical strategy is used, the society will have a specific form of money distribution in the final tick (here 9000). Therefore, the less variance the distribution has, the better resource allocation strategy has been used. Because it has more reduced the economic inequality among people (in terms of money distribution). Thus, the variance of money distribution in the society when it enters the final tick is another key factor for measuring the efficiency of the resource allocation strategy. The number of return periods of critical stages and variance of the money distribution in all runs have been presented for all of strategies in Table\ref{Application of strategy A up to 9000 ticks}, Table\ref{Application of strategy B up to 9000 ticks} and Table\ref{Application of strategy C up to 9000 ticks}.\\

  \begin{longtable}{|p{.12\textwidth} |p{.12\textwidth}| p{.12\textwidth} |p{.12\textwidth}|p{.12\textwidth}|p{.12\textwidth}| }
   \caption{Application of strategy A up to 9000 ticks}\label{Application of strategy A up to 9000 ticks}\\
   \hline
\# of runs & Mean &Variance &Total money of top 10\% &Total money of bottom 50\% &diff \\
 \hline
RUN 1 &2015&11328&11219&4465.974&-109\\
   \hline
RUN 2 &1517&10792&10888&4398.300&96\\
   \hline
RUN 3 &2287&11121&10784&4596.4128&-337\\
   \hline
RUN 4 &1386&10815&10650&4516.769&-165\\
   \hline
RUN 5 &2150&11001&0838&4489.615&-163\\
   \hline
RUN 6&1656&11208&11314&4341.547&106\\
   \hline
RUN 7 &2638&10745&10716&4479.010&-29\\
   \hline
RUN 8 &1590&11383&11298&4419.070&-85\\
   \hline
 RUN 9 &2368&10725&10364&4624.541&-361\\
   \hline
  RUN 10 &2059&10997&10727&4459.478&-270\\
   \hline
 \end{longtable} 
  \begin{longtable}{|p{.12\textwidth}|p{.12\textwidth}| p{.12\textwidth} |p{.12\textwidth}|p{.12\textwidth}|p{.12\textwidth}|}
   \caption{Application of strategy B up to 9000 ticks}\label{Application of strategy B up to 9000 ticks}\\
   \hline
\# of runs & Mean &Variance &Total money of top 10\% &Total money of bottom 50\% &diff \\
 \hline
RUN 1 &42&11032&11182&4399.110&150\\
   \hline
RUN 2 &47&11353&11459&4415.555&106\\
   \hline
RUN 3 &48&11475&11592&4443.134&117\\
   \hline
RUN 4 &18&11357&11474&4364.448&117\\
   \hline
RUN 5 &34&11256&11316&4367.595&60\\
   \hline
RUN 6&26 & 11594&11673&4378.737&79\\
   \hline
RUN 7 &37&11243&11351&4377.206&108\\
   \hline
RUN 8 &37&11213&11272&4426.685&59\\
   \hline
 RUN 9 &39&11351&11420&4435.002&69\\
   \hline
  RUN 10 &22&11252&11292&4379.559&40\\
   \hline
 \end{longtable} 
  \begin{longtable}{|p{.12\textwidth} | p{.12\textwidth} | p{.12\textwidth} |p{.12\textwidth}|p{.12\textwidth}|p{.12\textwidth}|}
   \caption{Application of strategy C up to 9000 ticks}\label{Application of strategy C up to 9000 ticks}\\
   \hline
\# of runs & Mean &Variance &Total money of top 10\% &Total money of bottom 50\% &diff \\
 \hline
RUN 1 &31&11224&11520.3&4302.669&296.33\\
   \hline
RUN 2 &19&11681&11760&4388.751&79\\
   \hline
RUN 3 &28&11464&11544.3&4409.166&80.33\\
   \hline
RUN 4 &14&11305&11565.3&4252.689&260.33\\
   \hline
RUN 5 &30&11371&11519.7&4315.761&148.66\\
   \hline
RUN 6&22&11667&11850&4394.670&182.99\\
   \hline
RUN 7 &24&11525&12082.7&4166.427&562.666\\
   \hline
RUN 8 &27&11579&11699&4363.750&119.99\\
   \hline
 RUN 9 &26&11286&11379.7&4326.632&93.66\\
   \hline
  RUN 10 &27&11635&11707.7&4311.794&72.66\\
   \hline
 \end{longtable}
 \begin{center}
\begin{figure}
\centering
    \includegraphics[scale = 0.8,trim = {0cm 0cm 0cm 0cm}, clip = true]{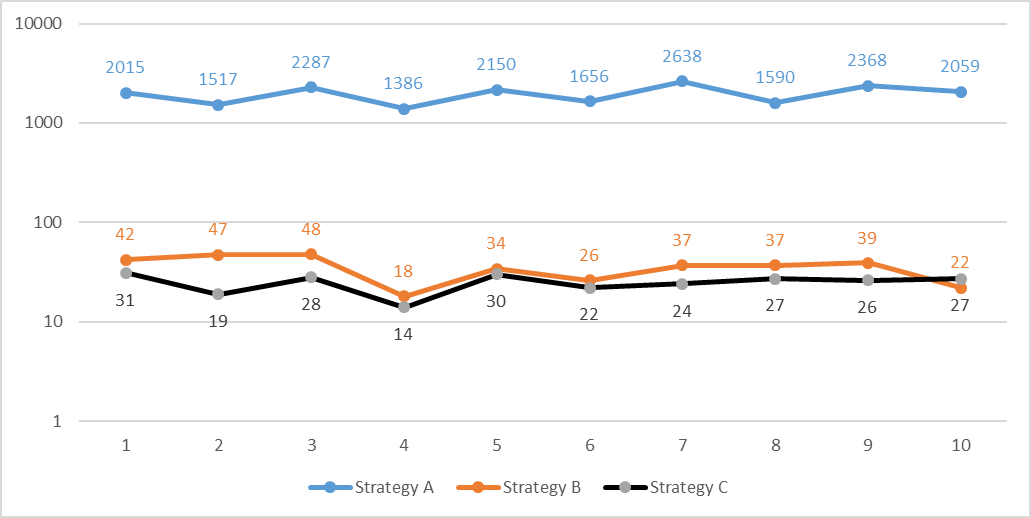}
    \caption{The sustainability of strategies}
    \label{sustainability}
\end{figure}
\end{center}
According to Table\ref{Application of strategy A up to 9000 ticks}, Table\ref{Application of strategy B up to 9000 ticks} and Table\ref{Application of strategy C up to 9000 ticks}, when strategy A is used, the average number of return periods is 1966.6 (roughly 1967) times Meaning that in all 10 runs, the economic system has returned to critical stage with the average of 1967 times. By applying strategy B, the system has shown a remarkably small average number of return periods which is 35. This number shows the second strategy is far more sustainable than strategy A in helping the system exit the critical stage for long time intervals. Strategy C has shown the number of 24.8 (roughly 25) for average of return periods in all runs.The number of return periods for each strategy has been visualized in Figure\ref{sustainability}: The sustainability of strategies. The vertical axis of this figure is the number of return periods that is made logarithmic (for better representation) and the horizontal axis stands for number of runs of simulation. As it can be seen, the strategy A has had the lowest level of sustainability while the strategy C has the highest one. Except run 10, strategy C has shown the lowest number of return periods for all other runs. Though this number doesn’t significantly differ from that of strategy B, it shows more sustainability.\\
 Another information that has be inferred from Table\ref{Application of strategy A up to 9000 ticks}, Table\ref{Application of strategy B up to 9000 ticks} and Table\ref{Application of strategy C up to 9000 ticks}, refers to how efficient each of allocation strategies has been in reducing the overall variance among people in terms of money distribution. This indicator shows how much agents differ from each other in terms of money volume.  When strategy A is applied, the average overall variance of money distribution is 4624.5 meaning that in all 10 runs of simulation, the average overall variance of distribution of money in economic system has been 4479.07. By executing strategy B, the average overall variance has decreased to 4398.70. While the charity paid in strategy B is 50 times more than that of strategy A (when systems enter critical stage), the average overall variance has not decreased remarkably (only 80.37 units). Strategy C has shown a better performance in reducing the variance of money distribution almost in all runs (except run 6). This strategy has had the overall average variance of 4323.2 for all 10 runs showing the fact that it is the most efficient strategy. The variance values of money distribution for each strategy have been visualized in Figure\ref{Variance}. The vertical axis of this figure is values of variance (for better representation) and the horizontal axis stands for number of runs of simulation. As it can be seen, the strategy C has had the lowest values of variances in all runs except run 6. Thus, this strategy is regarded to have the highest efficiency.
\begin{center}
\begin{figure}
\centering
    \includegraphics[scale = 0.8,trim = {0cm 0cm 0cm 0cm}, clip = true]{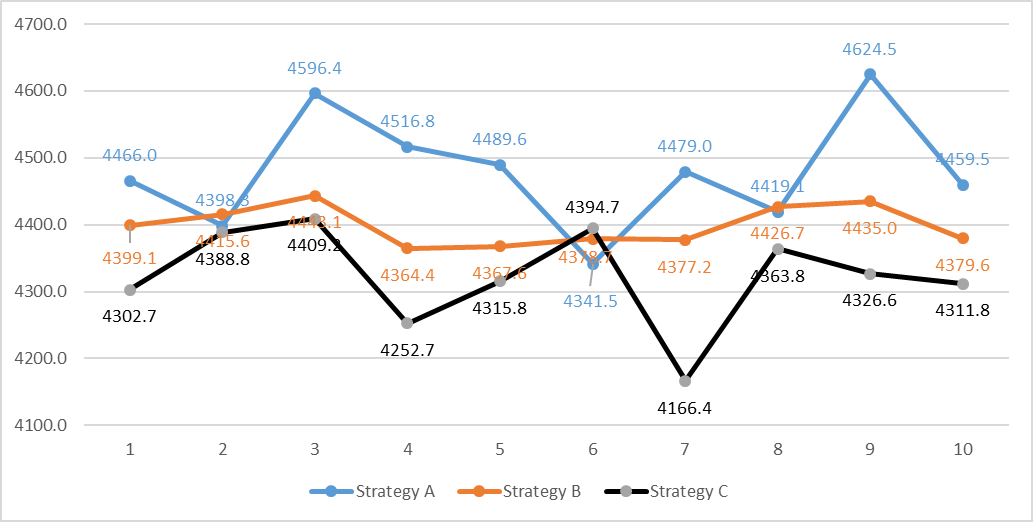}
    \caption{Variance}
    \label{Variance}
\end{figure}
\end{center}
\subsection{Remarks on Simulations }\label{section.simulations.Remarks}
These simulations have served two purposes. The first purpose was the explanation of how economic inequality emerges in an economic system. Results indicate that in a closed economic system when agents exchange the money, it tends to be accumulated in the hands of a very few number of agents over time. Therefore, the distribution of money in the society will follow a power law distribution. these results have found empirical support from the field of equilibrium statistical mechanics\cite{Dragulescu2000, Yakovenko2009}. As a fundamental law of this field, Boltzman-Gibbs law states that in a closed economic system, the total amount of money is conserved because it is not manufactured, consumed or destroyed so any conserved quantity in a big system should follow an exponential probability distribution. The second purpose was to simulate how charity and allocation strategies of charity entities can help reduce the economic inequality emerged in a system. The results showed that charity is highly effective in reducing the gap among economic deciles. So the more people pay charity the more they can decrease the gap. A part from charity, the results imply that the way charity entities allocate the money among lower economic deciles (i.e., bottom 50\%) is of paramount importance. In strategy A when system enters a critical stage, just one unit of money (as charity) is transited from the richest agent to poorest one. In strategy B, that amount of money becomes fifty times lager in each transition but the overall average variance (in comparison to average number of return periods) has not had a remarkable decrease. The money is made two times larger in strategy C but it shows a great decrease in overall average variance in contrast to that of strategy B. The main reason for this is not only because of money volume increase but also largely because of the ways money resources are taken from higher deciles and allocated to lower ones.
\section{ Conclusion} \label{section.conclusion}
Agent based modeling (ABM) has a high potentiality for modeling systems that are very hard or often impossible to capture by traditional modeling techniques such as PDEs, ODEs and even statistical modeling methods. In addition, ABM has shown a performance far better than a number of simulation methods such as discrete-event simulation (DES) and system dynamics (SD). A number of works have been published on this subject most often each of which has particularly dealt with one aspect of ABMs. For examples some works have only discussed the difference between ABM and EBM\cite{Sun2005,VanDykeParunak1998}. Some have only dealt with what of ABMs\cite{axtell2000agents,Chattoe-Brown2013,Epstein1999,Epstein1997,Heath2010a,Macy2002}and a number of works have just been conducted on issues of ABM verification and validation\cite{Law2008,NiaziMuazA;HussainAmir;Kolberg2017,PullumLauraL;Cui2012,Stone1974,Windrum2007,Xiang2005}, and replication and output analysis\cite{Axtell1996,Lee2015,wilensky2015introduction}. The major focus of this paper has been to help social sciences researchers (particularly economic planners) not only understand ABMs and gain a clear-cut big picture about them but also learn a step-by-step framework for developing them both systematically and rigorously.\\
 Like any scientific work, this work has had some limitations. The first limitation is the simplicity of the simulated economic systems. So extending the model to more elaborated levels can be a good subject for future studies. For example, human agents can be made more intelligent. They can have memory capacity, networked interactions, abilities for production, consumption and destruction, educational level, tendency for entrepreneurship and many other psychophysiological features. Moreover, new agent types can be added, for example, banks, factories, venture capital funds and so on. Such an extension can yield very interesting results. For instance, when factories hire more of their staff among those educated agents that belong to lower economic deciles, venture capital funds define some priorities for poor agents possessing high entrepreneurial tendency and a professional interaction is set among agents, some behavioral patterns will surprisingly emerge both at micro-scale (agent-level) and macro-scale (system-level).The second limitation is that just distribution of money has been the subject of this study and distribution of wealth or income has not been simulated. So, extending the system in order to show distribution of wealth and income will be a great contribution that can be made via future studies. The third limitation refers to the fact that in this paper, just two thought experiment examples have been provided and the validation process has not been done to full scale because of the space limitation (though it can be inferred via face validation). Therefore, applying the proposed framework for conducting real-world examples with empirical validations can be pursued in future studies. As the fourth limitation, the Netlogo toolkit has not been analyzed in details whereas it has some interestingly applied features for sensitivity analysis (i.e. BehaviorSpace), participatory simulation (i.e. Hubnet) and parameter regimes selection (i.e. Behaviorsearch). So having a specific study on Netlogo features and applications will be very useful for those social scientists who are eager to learn an easy but powerful ABM toolkit.
\bibliographystyle{ieeetran}
\bibliography{refs}
\end{document}